\documentclass[twocolumn]{aastex63}

\usepackage{paralist}

\usepackage{pifont}
\newcommand{\xmark}{\ding{53}}%

\newcommand{\TRISTAN}{{\bf \textsf{\small Tristan-MP}} }

\shorttitle{Cosmic ray ion vs lepton}
\shortauthors{Gupta, Caprioli \& Haggerty}
\date{\today}
\submitjournal{ApJ}
\graphicspath{{./}{figures/}}
\begin{document}

\title{Lepton-driven Non-resonant Streaming Instability}

\correspondingauthor{Siddhartha Gupta}
\email{gsiddhartha@uchicago.edu}

\author{Siddhartha Gupta$^{1}$, Damiano Caprioli}
\affiliation{Department of Astronomy and Astrophysics, University of Chicago, IL 60637, USA}
\author{Colby C. Haggerty}
\affiliation{Institute for Astronomy, University of Hawaii, Honolulu, HI, United States}

\begin{abstract}
A strong super-Alfv\'{e}nic drift of energetic particles (or cosmic rays, CRs) in a magnetized plasma can amplify the magnetic field significantly through non-resonant streaming instability (NRSI). 
While the traditional analysis is done for an ion current, here we use kinetic particle-in-cell simulations to study how the NRSI behaves when it is driven by electrons or by a mixture of electrons and positrons.
In particular, we characterize growth rate, spectrum, and helicity of the unstable modes, as well the level of magnetic field at saturation. 
Our results are potentially relevant for several space/astrophysical environments (e.g, electron strahl in the solar wind, at oblique non-relativistic shocks, around pulsar wind nebulae) and also in laboratory experiments.
\end{abstract}
%%%%%%%%%%%%%%%%%%%%%%%%%%%%%%%%%%%%%%%%%%%%%%%%%%%%%%%%%%%
%%%%%%%%%%%%%%%%%%%%%%%%%%%%%%%%%%%%%%%%%%%%%%%%%%%%%%%%%%%
%%%%%%%%%%%%%%%%%%%%%%%  Section 1  %%%%%%%%%%%%%%%%%%%%%%%
%%%%%%%%%%%%%%%%%%%%%%%%%%%%%%%%%%%%%%%%%%%%%%%%%%%%%%%%%%%
%%%%%%%%%%%%%%%%%%%%%%%%%%%%%%%%%%%%%%%%%%%%%%%%%%%%%%%%%%%
\section{Introduction} \label{sec:intro}
%%%%%%%%%%%%%%%%%%%%%%%%%%%%%%%%%%%%%%%%%%%%%%%%%%%%%%%%%%%
%%%%%%%%%%%%%%%%%%%%%%%%%%%%%%%%%%%%%%%%%%%%%%%%%%%%%%%%%%%
Interactions between energetic charged particles and a thermal background plasma generate a large variety of instabilities, ultimately fueled by the anisotropy of the non-thermal particles relative to the background plasma.
They are generally known as streaming instabilities (for reviews see, e.g., \citealt{Bykov+13,zweibel13}) and may produce large amplitude modes over a broad range of scales, from the ion-skin depth ($\sim 100$ km in the interstellar medium) to the pc-scale of the gyroradius of the highest-energy Galactic cosmic rays (CRs). 
These instabilities are crucial for the generation of magnetic fields, the acceleration and propagation of non-thermal particles, and for the heating of space and astrophysical plasmas. 
Finally, modern laser facilities are unlocking the possibility to study streaming instabilities also in laboratory, even in the collisionless regime \citep[e.g.,][]{jao+19}. 

In the context of the interactions between CRs and a thermal background plasma, there are two main regimes of interest: the resonant and non-resonant streaming instabilities (hereafter RSI and NRSI, respectively), with the latter dominating for strong CR currents  \citep{bell04,amato+09}.

The NRSI is characterized by a fastest-growing mode with wavelength $\lambda_{\rm fast} \approx c\,B_{\rm 0} /J_{\rm cr}$, where $J_{\rm cr}$ is the CR current density in the direction of the mean magnetic field $B_0$ and $c$ is the speed of light.
The instability is dubbed non-resonant because the wavelength of the fastest growing mode is shorter than $R_{\rm L}$, the CR gyroradius, i.e., $\lambda_{\rm fast} \ll R_{\rm L}$ and the unstable modes have right-handed circular polarization (so the fastest growing mode of magnetic field does not rotate in the same direction as current-carrying CR ions).

While a magnetized plasma is typically considered for the NRSI, it is worth mentioning that this instability can be triggered even in the absence of initial magnetic field due to the results of other interactions such as the Weibel instability \citep{weibel59}, which can provide the seed magnetic field \citep[see, e.g.,][]{peterson+21}.

Thought formally present already in the derivations of \citet{achterberg83,winske+84}, non-resonant modes were recognized as crucial for CR scattering by \citet{bell04}, after it has been shown that CR-driven instabilities may strongly amplify the initial magnetic field to non-linear values of $\delta B/B_0\gg 1$ \citep{lucek+00,bell+01}.

\subsection{Lepton-driven NRSI}
The NRSI (also called non-resonant hybrid, or simply Bell, instability) has been studied extensively  with analytical, MHD, and kinetic approaches \citep[e.g.,][]{niemiec+08, gargate+10, zirakashvili+08,amato+09,riquelme+09, reville+13,matthews+17,   Weidl+19a,haggerty+19p,zacharegkas+19p,marret+21}, always under the assumption that the current is carried by protons.
The motivation for this choice is that the electron/ion ratio in CR fluxes at Earth is rather small $\lesssim 10^{-2}-10^{-3}$, as it is in sources such as supernova remnants \citep[e.g.,][]{berezhko+04a,morlino+12}.

Nevertheless, there are several instances in which a strong current driven by non-thermal leptons may arise.
For instance, in quasi-perpendicular shocks (where the pre-shock magnetic field makes an angle $\gtrsim 60\deg$ with the shock normal) the injection of thermal ions is suppressed \citep[][]{caprioli+15} but electrons can still be injected and undergo shock acceleration \citep[e.g.,][]{guo+14a,guo+14b, xu+20, bohdan+19a}.
Another environment where strong lepton currents can be generated are pulsar wind nebulae (PWNe), which are leptonic sources that can accelerate electrons and positrons up to PeV energies. 
Recently, $\gamma$-ray halos have been discovered around nearby PWNe \citep{hawc17,schroer+21}, attesting to the fact that escaping leptons can strongly modify the interstellar magnetic fields, leading to particle self-confinement. 
Note that, even if the seeds for PWN relativistic particles are likely magnetospheric pairs, the highest-energy leptons are found to be of a given sign, depending on the relative orientation of the pulsar magnetic and rotation axes \citep[e.g.,][]{cerutti+15,philippov17, philippov+18}.

There are also plasma systems closer to Earth where non-thermal electrons are important, such as the strahl in the solar wind or planetary bow shocks \citep[e.g.,][]{malaspina+20,masters+13,wilson+16,masters+17}.
Within 30 $R_\odot$ of the sun, the momentum flux of the electron strahl is within an order of magnitude of reaching the non-resonant threshold, that will be discussed in this work \citep[as determined from recent in situ measurements reported from the first few perihelion passes of Parker Solar Probe]{kasper+19,halekas+19}. 
The nearest to the instability threshold suggests that the electron driven NRSI may be occurring closer to the sun where the momentum flux of non-thermal electrons is expected to be larger, and that this instability can be responsible for the scattering of the strahl.

Electron-driven NRSI may finally be of interest for laboratory plasma experiments \citep[][]{bret+10}. 
With very powerful lasers, it is possible to reproduce the collisionless conditions typical of astrophysical systems.
While experiments have not been able to recreate the condition to drive the NRSI with ions, to our knowledge, a few works have attempted to do so with electrons (e.g., \citealt{jao+19}). 
Therefore, it is important to put forward a theory of lepton-driven NRSI and validate it via kinetic simulations, which is the scope of this work.

Bell's derivation of the NRSI \citep{bell04} highlights how, as long as the CRs are infinitely rigid, the maximally unstable mode and its associated growth rate depend on the compensating current induced in the background plasma.
At the first order in the small parameter $n_{\rm cr}/n$, i.e., the ratio in CR to thermal number density, the NRSI growth rate is independent of the composition of the CR distribution and only depends on the net induced return current \citep[also see][]{amato+09,Weidl+19a};
however, it is non trivial that the return current, which is supported by the light thermal electrons, behaves the same for negatively-charged CRs, or for CR distributions with both positive and negative charges.

In this work, we derive the NRSI for CRs with arbitrary mass and charge and in particular to address the following questions:
{\begin{compactitem}
\setlength\itemsep{0.1em}
\item What are the necessary conditions for having lepton-driven NRSI?
\item What are the properties of the fastest growing modes (polarization, wavelength, growth rate)?
\item Can CRs with a mixed (e.g., electrons and positrons) composition produce NRSI?
\item Is the saturation of the amplified magnetic field the same as in the ion-driven case?
\end{compactitem}}

We begin by outlining the analytical linear theory for the NRSI driven by CRs of arbitrary mass and charge in \S\ref{sec:analytics}. 
In \S\ref{sec:setup} we introduce self-consistent particle-in-cell (PIC) simulations used to test both ion- and electron-driven NRSI (henceforth, CR-I and CR-E), which are compared and discussed in \S\ref{sec:results}. 
The implications of this study to different plasma backgrounds (e.g., electron-positron) and mixed compositions of CRs are also discussed in \S\ref{sec:results}. 
We conclude in \S\ref{sec:summary}.

%%%%%%%%%%%%%%%%%%%%%%%%%%%%%%%%%%%%%%%%%%%%%%%%%%%%%%%%%%%
%%%%%%%%%%%%%%%%%%%%%%%%%%%%%%%%%%%%%%%%%%%%%%%%%%%%%%%%%%%
%%%%%%%%%%%%%%%%%%%%%%%  Section 2  %%%%%%%%%%%%%%%%%%%%%%%
%%%%%%%%%%%%%%%%%%%%%%%%%%%%%%%%%%%%%%%%%%%%%%%%%%%%%%%%%%%
%%%%%%%%%%%%%%%%%%%%%%%%%%%%%%%%%%%%%%%%%%%%%%%%%%%%%%%%%%%
\section{Linear theory}\label{sec:analytics}
%%%%%%%%%%%%%%%%%%%%%%%%%%%%%%%%%%%%%%%%%%%%%%%%%%%%%%%%%%%
%%%%%%%%%%%%%%%%%%%%%%%%%%%%%%%%%%%%%%%%%%%%%%%%%%%%%%%%%%%
%%%%%%%%%%%%%%%%%%%%%%%%%%%%%%%%%%%%%%%%%%%%%%%%%%
\begin{figure*}
\centering
\begin{minipage}{0.4\linewidth}
\includegraphics[height=2.2in,width=3.in]{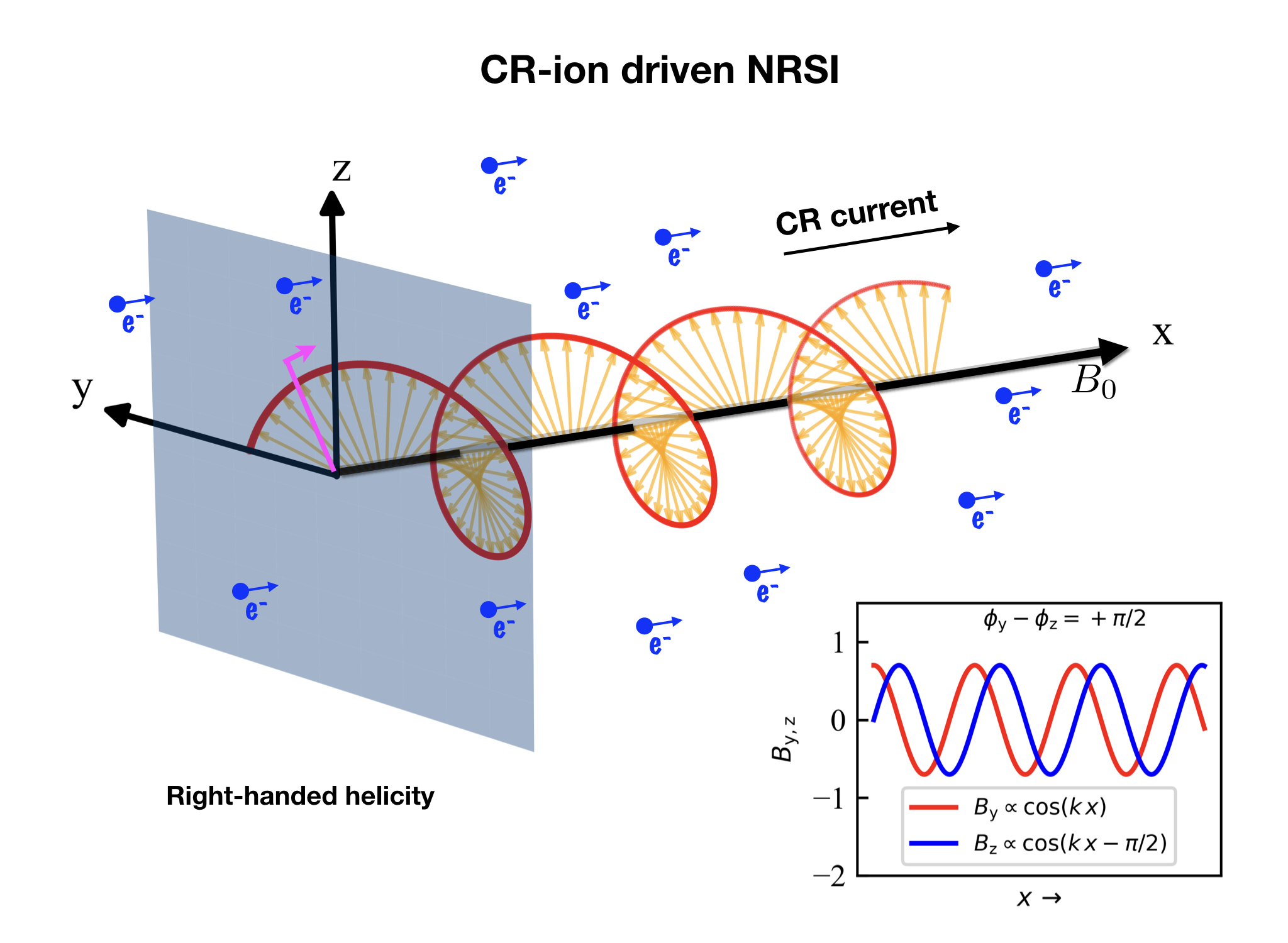}
\end{minipage}
\hspace{0.1\linewidth}
\begin{minipage}{0.4\linewidth}
\centering
\includegraphics[height=2.2in,width=3.in]{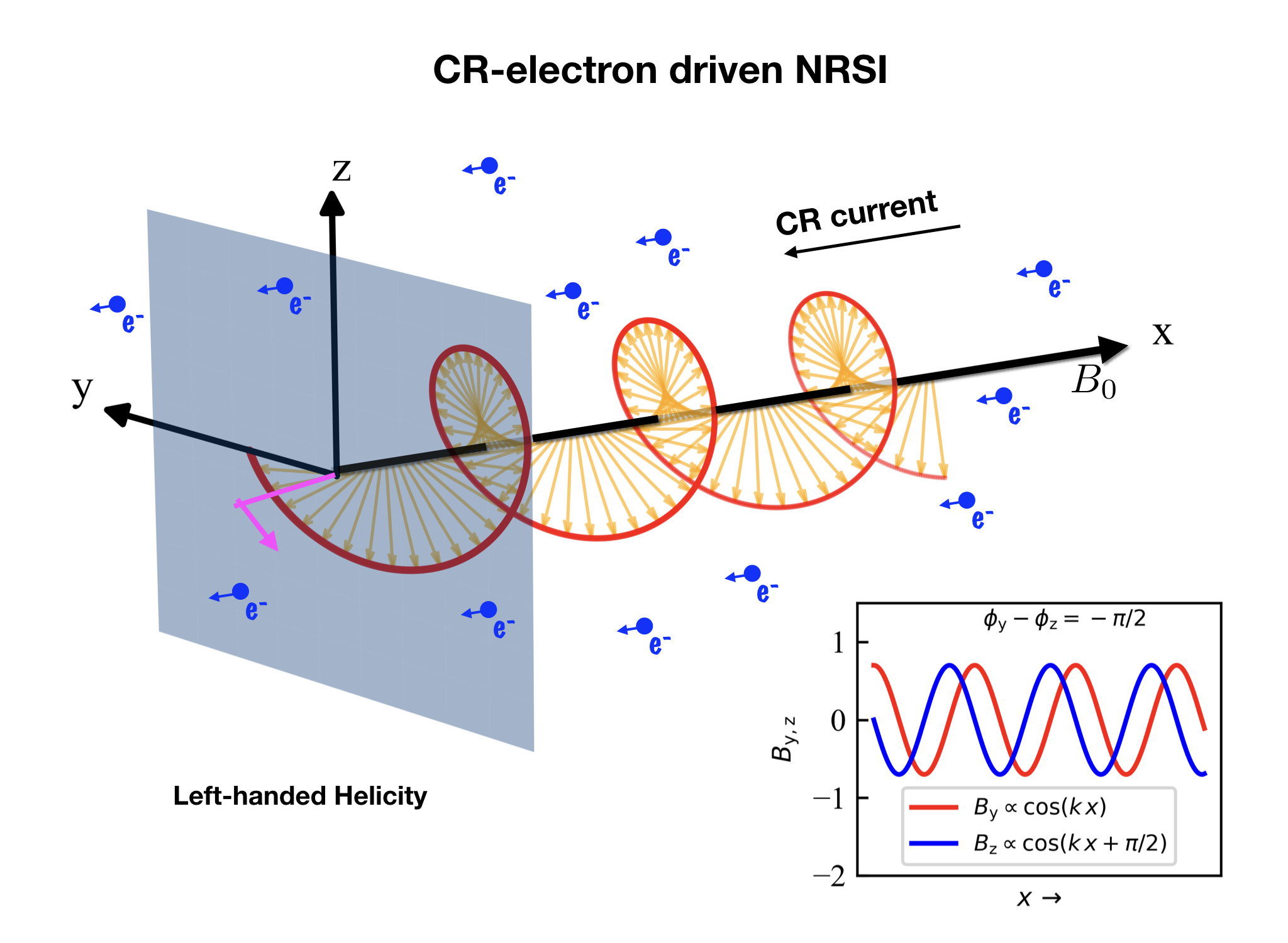}
\end{minipage}
\caption{Schematic diagram showing the structure of the amplified magnetic field for CR-ion (left) and electron (right) driven NRSI. CRs are drifting parallel to an initial magnetic field $B_{\rm 0}\hat{x}$, which produces a current denoted by a black arrow. Blue circles represent the plasma electrons, where the arrows indicate the drift velocity (in the plasma frame) that compensates the CR current. This system is unstable and produces transverse ($y,z$ directions) waves. The expected phase difference between the transverse (growing) components of the magnetic field (${ B_{\rm y,z}}$) is shown in the box. Considering the transverse components $B_{\rm y,z}\propto \exp[j(kx+\phi_{\rm y,z})]$, for a given $k$, if $\Delta \phi = \phi_{\rm y}-\phi_{\rm z} > 0$ then transverse B-field, $B_{\rm y} {\bf \hat{y}}+B_{\rm z}{\bf \hat{z}}$, rotates clockwise as one moves along the positive x-axis. We refer this as a right-handed mode and opposite to it as a left handed mode. The figure shows that in CR-ion (electron) case, the waves are right (left) handed with respect to the direction of the initial magnetic field ($B_{\rm 0}$).}\label{fig:schdia}
\end{figure*}
%%%%%%%%%%%%%%%%%%%%%%%%%%%%%%%%%%%%%%%%%%%%%%%%%%
The theory of NRSI driven by energetic CRs propagating along magnetic field lines has been studied in both the fluid and kinetic limits \citep{bell04,amato+09,riquelme+09, zweibel+10};
here, we present a simple derivation, which explicitly assumes that resonant interactions between CRs and growing waves are negligible (see \S \ref{subsec:kl-limit}), for an arbitrary mass and charge of CRs.

The bulk motion of CRs produces a strong current in the plasma, which needs to be compensated by the drift of thermal background electrons.
Such a drift velocity can be found by balancing the currents of the CRs and the background, i.e.: 
\begin{equation}\label{eq:veloe}
{\bf v}_{\rm e} = s_{\rm cr}\frac{n_{\rm cr}}{n_{\rm e}} {\bf v}_{\rm d} \ .
\end{equation}
Here ${\bf v}_{\rm d}$ is the CR drift velocity with respect to the thermal ions (the analysis is done in the ion rest-frame), and  $n_{\rm cr}$ and $n_{\rm e}$ are the number density of CRs and background electrons, respectively.
We pose ${\bf v}_{\rm d}=v_{\rm d} \hat{\bf x}$, so that the return current electrons drift along the positive/negative $x$-axis, depending on the sign of the charge of the CRs ($s_{\rm cr}=\pm 1$), as sketched in Figure \ref{fig:schdia}. 
Quasi-neutrality requires that the number density of ions, electrons and CRs must balance, i.e., $n_{\rm e} = n_{\rm i} +  s_{\rm cr}\,n_{\rm cr}$.
In typical astrophysical applications, the CR number density is much smaller than the density of the background plasma ($n_{\rm cr} \ll n_i \approx n_{\rm e} \equiv n_{\rm 0}$), so that $v_{\rm e} \ll v_{\rm d}$.

The motion of any particle in the species $\alpha$ is given by the Lorentz force:
\begin{equation}\label{eq:nonrel_motion}
m_{\rm \alpha}\frac{\partial {\bf v}_{\rm \alpha}}{\partial t} =q_{\rm \alpha}\left[{\bf E}+\frac{{\bf v_{\rm \alpha}}}{c}\times {\bf B}\right]\, ,
\end{equation}
where ${\bf v_{\rm \alpha}}$ is the velocity of a particle of mass $m_{\rm \alpha}$ and charge $q_{\rm \alpha}$ (representing ions, electrons, hereafter $\alpha = i,e$), ${\bf E}$ and ${\bf B}$ are the electric and magnetic field. 
We consider a system with no initial electric field (${\bf E} = 0$) and a uniform magnetic field ${\bf B}=B_{\rm 0}\hat{\bf x}$.
Assuming that the background population is sufficiently cold, so that initially ${\bf v_{\rm i}}\approx 0$ and ${\bf v_{\rm e}}$ is given by Equation \ref{eq:veloe}, we can linearize Equation \ref{eq:nonrel_motion} along with Maxwell equations (for details see Appendix \ref{app:disper}) by considering small plane-wave perturbations $\propto\exp[{j(k x-\omega t)}]$ \citep[][]{krall-trivel,achterberg83,choudhuri98}, where $k$ and $\omega$ are the usual (parallel) wavenumber and the angular frequency of the plasma modes. 
With an additional assumption that $|\omega|\ll \omega_{\rm ci}$, i.e., that both the instability growth rate (the imaginary part of $\omega$) and the phase speed (the real part of $\omega$) of the modes are much smaller than the ion cyclotron frequency, $\omega_{\rm ci}$, we obtain the following dispersion relations for left- and right-handed (LH, RH) modes \footnote{The convention is illustrated in Figure \ref{fig:schdia}.}:
\begin{equation} \label{eq:disp}
\omega_{\rm R,L} \approx 
%\pm s_{\rm cr}\frac{v_{\rm A0}^2\,k_{\rm u}}{2\,v_{\rm d}} 
\pm s_{\rm cr} \frac{1}{2}\frac{n_{\rm cr}}{n_{\rm e}}\, \omega_{\rm ci}
+ v_{\rm A0}\,k\,\left[1\mp\,s_{\rm cr}\,\frac{k_{\rm u}}{k}\right]^{1/2} .
\end{equation}
Here $v_{\rm A0} \equiv B_{\rm 0}/\left(4\pi\,m_{\rm i}\, n_{\rm 0}\right)^{1/2}$ is the Alfv\'{e}n speed and we have introduced the critical wavenumber
\begin{equation}
     k_{\rm u} \equiv \frac{\omega_{\rm ci}\, |v_{\rm e}|}{v_{\rm A0}^2} = 
     \frac{\omega_{\rm ci}}{v_{\rm A0}}\frac{n_{\rm cr}}{n_{\rm e}}\frac{v_{\rm d}}{v_{\rm A0}} .
\end{equation}

\noindent
This makes it evident that, for a given CR charge $s_{\rm cr}$, one branch of modes becomes unstable for $k<k_{\rm u}$ and for small $k$, the growth rate is suppressed $\propto k^{1/2}$. 
The phase difference between transverse components of the perturbed magnetic field is (see Equations \ref{eq:MFeq} and \ref{eq:alphabeta}): 
\begin{equation}\label{eq:disp_0}
%\delta 
\Delta \phi\equiv \phi_{\rm y}-\phi_{\rm z} = \pm\, \frac{\pi}{2} \, .
\end{equation}
Therefore, the helicity of the transverse magnetic field is determined by the upper/lower sign of the dispersion relation (Equation \ref{eq:disp}), with the positive and negative sign corresponding to R-handed  and L-handed modes, respectively.
Figure \ref{fig:schdia} sketches the expected helicity of the resulting modes for CR-I and CR-E driven cases and Figure \ref{fig:diffregime} summarizes the different regimes of Equation \ref{eq:disp} as a function of $k/k_{\rm u}$.

\subsection{${\rm \bf Regime\,I:}\, k > k_{\rm u}$}\label{subsec:ku-limit}
\begin{figure}
\centering
\includegraphics[width=0.47\textwidth]{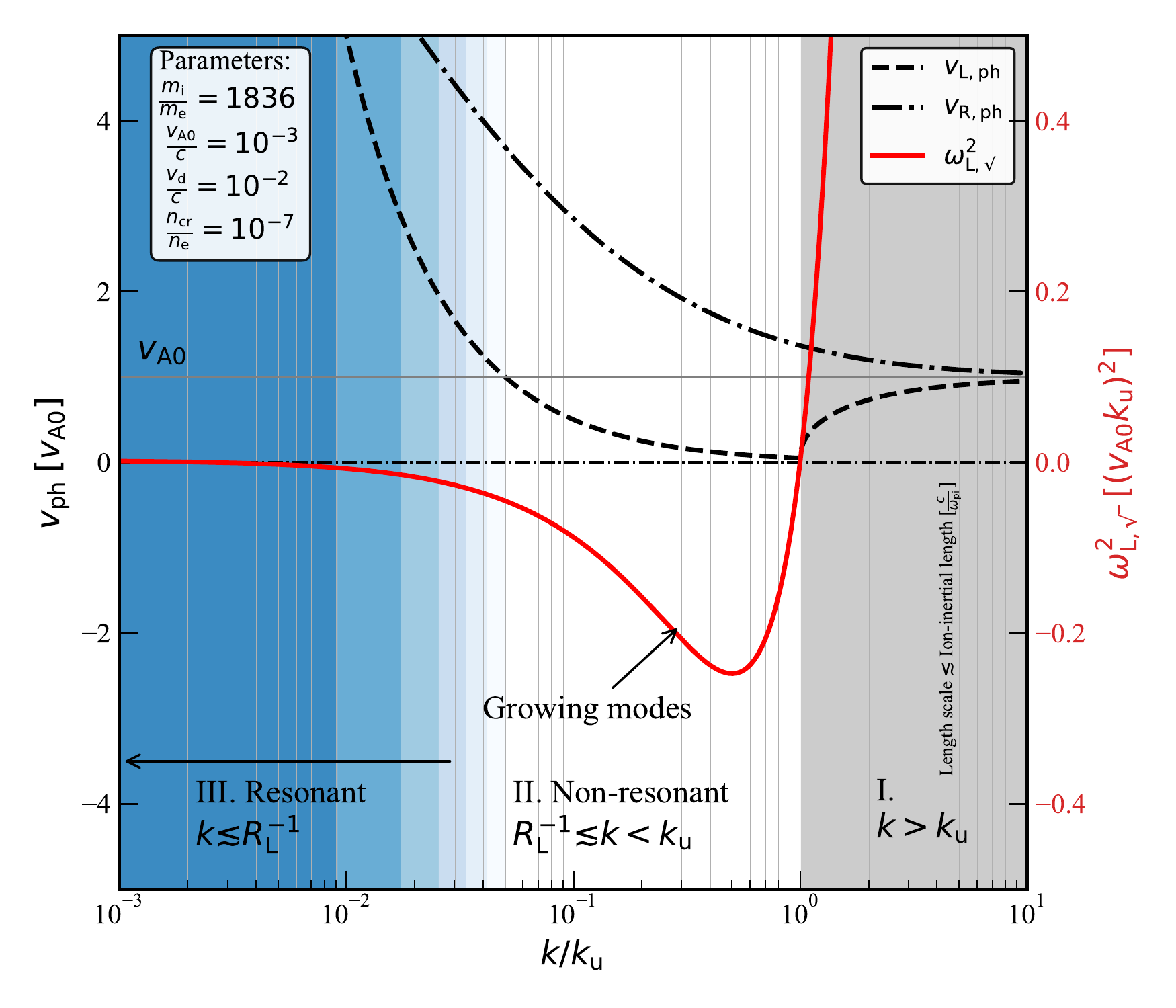}
\caption{Phase speed and the growing/damping part of the angular frequency (the second term in the right side of Equation \ref{eq:disp}: $\omega^2_{\rm L, \sqrt{\,}}$) as a function of $k$ for electron-driven NRSI (using $s_{\rm cr}=-1$ in Equation \ref{eq:disp}). The figure shows $\omega^2_{\rm L, \sqrt{\,}} <0$ 
%\dam{Replace the square root notation.}
when $\kappa < k_{\rm u}$ (Regime  II). In this regime, although the phase speed of the LH modes (dashed curve) are smaller than both the CR drift velocity ($v_{\rm d}= 10\,v_{\rm A0}$) and the Alfv\'{e}n speed (in the limit $k\rightarrow k_{\rm u}$), the waves gain energy. This is the non-resonant mode (\citealt{bell04}).
}\label{fig:diffregime}
\end{figure}

%\subsection{ {\bf Regime I} : $k > k_{\rm u}$}\label{subsec:ku-limit}
%%%%%%%%%%%%%%%%%%%%%%%%%%%%%%%%%%%%%%%%%%%%%%%%%%%%%%%%%%%
This regime (gray-shaded region in Figure \ref{fig:diffregime}) corresponds to oscillatory modes with wavelength smaller than the ion inertial length ($c/\omega_{\rm pi}$; $\omega_{\rm pi}$ the plasma frequency for ions).

%%%%%%%%%%%%%%%%%%%%%%%%%%%%%%%%%%%%%%%%%%%%%%%%%%%%%%%%%%%
%
\subsection{${\rm \bf Regime\,II:}\, k< k_{\rm u}$}\label{subsec:nonreso-limit}
%%%%%%%%%%%%%%%%%%%%%%%%%%%%%%%%%%%%%%%%%%%%%%%%%%%%%%%%%%%
In this regime (white region in Figure \ref{fig:diffregime}), $\omega$ has both real and imaginary parts. 
Depending on the CR charge, either RH or LH modes will be amplified:
for CR-I/CR-E (i.e., $s_{\rm cr}=+1/-1$) waves grow when the upper/lower sign of Equation \ref{eq:disp} is chosen, corresponding to RH and LH modes, respectively.
While Equation \ref{eq:disp} accurately captures the growth rate of the most unstable branch in the limit $v_{\rm d}\gg v_{\rm A}$, the present derivation does not extend to the weak-current limit, in which resonant modes grow with much smaller rate;
the RSI solution appear only in a kinetic calculation done in the proper wave frame \citep{zweibel79,achterberg83, bell04,amato+09}.

From Equation \ref{eq:disp} we also see that the phase speed of the growing modes (RH/LH in CR-I/CR-E case) is
\begin{equation}
    v_{\rm ph}\approx \frac{1}{2}\frac{n_{\rm cr}}{n_{\rm e}}\frac{\omega_{\rm ci}}{k}=\frac{1}{2}\frac{v^2_{\rm A0}}{v_{\rm d}}\frac{k_{\rm u}}{k},
\end{equation}
consistent with \citet{riquelme+09}. 
Since we have taken $v_{\rm A0} \ll v_{\rm d}$, the phase velocity (dashed curve in Figure \ref{fig:diffregime}) and group velocity are much smaller than the drift velocity of plasma electrons, i.e., non-resonant modes are almost stationary in the plasma frame as $k\rightarrow k_{\rm u}$. Whereas, the phase speed of the other branch (dash-dotted curve in Figure \ref{fig:diffregime}) is typically larger than $v_{\rm A0}$; 
for a smaller $k/k_{\rm u}$, close to resonant scales, $v_{\rm ph}$ of both LH and RH branches is larger than $v_{\rm A0}$. 
This can be important in determining the speed of the CR scattering centers in shock environments, where they contribute in shaping the shock dynamics and the CR spectra \citep[e.g.,][]{haggerty+20,caprioli+20}. 

It is straightforward to show that (Appendix \ref{app:disper}), irrespective of the composition of CRs, the fastest-growing mode is at $k_{\rm fast}=k_{\rm u}/2$:
\begin{equation}\label{eq:kmax}
k_{\rm fast} \equiv \frac{1}{2} \frac{v_{\rm e}}{v_{\rm A0}}\,\frac{1}{d_{\rm i}}=\frac{1}{2} \frac{n_{\rm cr}}{n_{\rm e}}\frac{v_{\rm d}}{v_{\rm A0}} \frac{1}{d_{\rm i}},
\end{equation}
where $d_{\rm i}= c/\omega_{\rm pi}=v_{\rm A0}/\omega_{\rm ci}$ is the ion skin depth, and the corresponding growth rate is 
\begin{equation}\label{eq:wmax}
\gamma_{\rm fast} \equiv k_{\rm fast}\, v_{\rm A0} = \frac{1}{2}\frac{n_{\rm cr}}{n_{\rm e}}\frac{v_{\rm d}}{v_{\rm A0}}\,\omega_{\rm ci}.
\end{equation}
%%%%%%%%%%%%%%%%%%%%%%%%%%%%%%%%%%%%%%%%%%%%%%%%
\begin{table*}
\begin{center}
\begin{tabular}{l | c c c c c  | c c c | c  c c c }
   \hline\hline
   Run  & $N_{\rm x}$& $N_{\rm y}$&  $\frac{d_{\rm e}}{\Delta_{\rm x}}$&  $\frac{m_{\rm i} }{m_{\rm e}}$ &  $\frac{v_{\rm A0} }{c}$  & $\frac{v_{\rm bst}}{c}$ & $\frac{n_{\rm cr}}{n_{0}}$ & $\frac{p^{\prime}_{\rm cr}}{m_{\rm i}\,c}$ & $\frac{v_{\rm d}}{c}$  & $\xi$   & $\frac{k_{\rm fast}}{1/d_{\rm e}}$ & $\frac{\gamma_{\rm fast}}{\omega_{\rm pe}}$ \\ 
    &  &  & & & $\times 10^{-2}$ &  & $\times 10^{-3}$ &  &  & &   $\times 10^{-2}$ & $\times 10^{-4}$ \\ 
    \hline
    A. EI-S-$\xi340$\,$^{\bigstar}$ & $3\times 10^4$ &  $5$  &  $5$ &  $100$ &  $1$ & $0.8$ & $4$ & $10$  &   $0.635$  & $340$ & $1.27$  & $1.27$\\
    B. EI-S-$\xi56$ & $3\times 10^4$ & $5$ &  $5$ &  $100$ &$1$ & $0.8$ & $4$ & $1$   &    $0.740$ & $56$  & $1.48$  & $1.48$ \\
    C. EI-S-$\xi21$ & $5\times 10^4$ &  $5$ & $5$ & $100$  & $4$ &  $0.8$ & $10$ & $10$     &     $0.635$ & $21$  & $0.79$  & $3.2$\\
    D. EI-S-$\xi11$ & $10\times 10^4$ & $5$ & $5$ & $100$  & $1$ &  $0.2$ & $4$ & $10$     &     $0.135$ & $11$& $0.27$  & $0.27$\\
    \hline
    E. EI-M-$\xi170$  & $3\times 10^4$ &$5$ &  $5$ & $100$  & $1$ &  $0.8$ & $6,4$ & $10,10$     &     $0.635$ & $170$& $0.635$  & $0.635$\\
    F. EI-M-$\xi68$  & $6\times 10^4$ &$5$ &  $5$ & $100$  & $1$ &  $0.8$ & $4.8,4$ & $10,10$     &     $0.635$ & $68$& $0.254$  & $0.254$\\
    G. EI-M-$\xi0$  & $500$ &$500$ &  $5$ & $100$  & $1$ &  $0.8$ & $4,4$ & $10,10$     &     $0.635$ & $0$ & \xmark  & \xmark\\
    \hline
    H. EP-S-$\xi42$ & $3\times 10^4$ &  $5$ &  $5$ & $1$ &   $1$ &  $0.8$ & $1$ & $10$     &     $0.635$ & $42$& $2.25$  & $1.59$\\
    I. EP-M-$\xi0$ & $4000$ &  $250$ &  $5$  & $1$ &   $0.32$ &  $0.8$ & $1,1$ & $10,10$     &     $0.635$ & $0$ & \xmark  & \xmark\\
    
    \hline
\end{tabular}
\end{center}
\centering
\caption{Simulation parameters for different runs. Columns indicate: number of cells along the $x$ and $y$ directions, number of cells per electron skin depth, mass ratio, Alfv\'en speed, boost speed, CR density and momentum in their rest frame, effective drift speed, parameter $\xi$ (Equation \ref{eq:nrlimit}), and expected $k$-mode and growth rate for the most unstable mode (Equations \ref{eq:kmax} and \ref{eq:wmax}). 
The nomenclature `EI-S-$\xi21$' represents a run where the background is made of electron-ion (EI) plasma, CR beam contains a single (S) charged species, and $\xi\approx 21$.
Runs E--G: CRs contain a mixed (M) population of positive and negative charges where $m_{\rm i}\gamma_{\rm i}=m_{\rm e}\gamma_{\rm e}=10$ (pair beam in an electron-ion plasma). 
Runs H and I are similar to the previous case, except that here $m_{\rm i}=m_{\rm e}$ (pair beam in pair plasma). The symbol `${\bigstar}$' marks the benchmark simulation.}\label{tab:simpara}
\end{table*}
%%%%%%%%%%%%%%%%%%%%%%%%%%%%%%%%%%%%%%%%%%%%%%%%
%%%%%%%%%%%%%%%%%%%%%%%%%%%%%%%%%%%%%%%%%%%%%%%%%%%%%%%%%%%
%
\subsection{${\rm \bf Regime\,III:}\, k\ll k_{\rm u}$}\label{subsec:kl-limit}
%%%%%%%%%%%%%%%%%%%%%%%%%%%%%%%%%%%%%%%%%%%%%%%%%%%%%%%%%%%
The above derivation is oblivious to any resonant interaction between CRs and growing modes, and hence holds as long as $k R_{\rm L}\gg 1$, where $R_{\rm L}=p_{\rm cr}c/eB_0$ is the gyroradius of a CR with momentum $p_{\rm cr}$;
such an assumption must break for sufficiently small $k$  (blue-shaded region III in Figure \ref{fig:diffregime}). 
Fully kinetic calculations show that in this regime the NRSI becomes comparable to, or even less important than, the RSI \citep{amato+09,haggerty+19p}. 
Although the exact transition from NRSI to RSI depends on the shape of the CR distribution function, in general the NRSI dominates when $k_{\rm fast} R_{\rm L}\gg 1$, which corresponds to:
\begin{equation} \label{eq:nrlimit}
\xi\equiv \frac{n_{\rm cr}}{n_{\rm 0}}\,\frac{p_{\rm cr}}{m_{\rm i}}  \frac{v_{\rm d}}{v_{\rm A0}^2}
\equiv \frac{1}{2}\frac{P_{\rm cr}}{P_{\rm B0}}
\gg 1,
\end{equation}
where $P_{\rm B0}$ is the magnetic pressure and $P_{\rm cr}$ is the CR momentum flux (anisotropic pressure) along $x$. 
In general, the NRSI can be triggered if a charged species has an anisotropic pressure that exceeds the magnetic one; to some extent, it could be thought of as a firehose instability driven by charged particles \citep[e.g.,][and references therein]{shapiro+98}.

Note that $\xi$ in  Equation \ref{eq:nrlimit} depends on the momentum of CR particles divided by the \emph{ion} mass, which means that for relativistic electrons to satisfy the condition $\xi\gg 1$, their Lorentz factor $\gamma_{\rm e}$ has to be a factor of $m_{\rm i}/m_{\rm e}\sim 2000$ larger than for the canonical ion-driven NRSI. 

When leptons with large Lorentz factors are involved, it is worth checking the condition that the NRSI growth rate is larger than the synchrotron loss rate \citep{rybiki+86}.
Losses are negligible\footnote{Technically, for large values of $\xi$, when $\delta B\gg B_0$ is expected, losses may affect the NRSI saturation for smaller values of $\gamma_e$.} as long as the electron Lorentz factor $\gamma_e$ satisfies:
\begin{equation}\label{eq:gam_upper}
    %\gamma_{\rm cr} \ll 4.5\pi \frac{n_{\rm cr}}{n_{\rm 0}} \frac{v_{\rm d}c}{v_{\rm A0}^2}\left(\frac{m_{\rm cr}}{m_{\rm i}}\right)^3 n_{\rm0}\, d_{\rm i}^3\, .
    \gamma_{\rm e} \ll 3.7\times10^{12}\, \frac{n_{\rm cr}}{n_{\rm 0}} \frac{v_{\rm d}}{v_{\rm A0}} \left(\frac{B_{\rm 0}}{G}\right)^{-1}.
\end{equation}
In astrophysical environments, e.g., for shocks in the interstellar medium, one has $n_{\rm cr}/n_{\rm 0}\sim 10^{-7}$, and $B_{\rm 0}\sim 3\, \mu$G, $v_{\rm d}/v_{\rm A0}\sim 10$, for which Equation \ref{eq:gam_upper} returns an upper limit of $\gamma_{\rm e}\sim 10^{12}$, i.e., the effect of synchrotron losses are negligible. 
However, in laboratory experiments the above condition must be reckoned with, since $B_{\rm 0}\sim 10^3$ G, $n_0$ is large, which are needed for satisfying $\gamma_{\rm fast}< \omega_{\rm ci}$ \citep{jao+19}.

Combining Equations \ref{eq:wmax} and \ref{eq:nrlimit}, and the condition $\gamma_{\rm fast} < \omega_{\rm ci}$, we find that the necessary momentum to drive the NRSI by an arbitrary mass of CRs is
\begin{equation}\label{eq:pcrmin}
p_{\rm cr} \gg m_{\rm i} v_{\rm A0}\frac{\omega_{\rm ci}}{\gamma_{\rm fast}} \ . 
\end{equation}
In the following sections we test these expectations with self-consistent kinetic simulations using CR beams with species of different mass and charges.
%%%%%%%%%%%%%%%%%%%%%%%%%%%%%%%%%%%%%%%%%%%%%%%%%%%%%%%%%%%
%%%%%%%%%%%%%%%%%%%%%%%%%%%%%%%%%%%%%%%%%%%%%%%%%%%%%%%%%%%
%%%%%%%%%%%%%%%%%%%%%%%  Section 3  %%%%%%%%%%%%%%%%%%%%%%%
%%%%%%%%%%%%%%%%%%%%%%%%%%%%%%%%%%%%%%%%%%%%%%%%%%%%%%%%%%%
%%%%%%%%%%%%%%%%%%%%%%%%%%%%%%%%%%%%%%%%%%%%%%%%%%%%%%%%%%%
\section{Numerical setup}\label{sec:setup}
%%%%%%%%%%%%%%%%%%%%%%%%%%%%%%%%%%%%%%%%%%%%%%%%%%%%%%%%%%%
%%%%%%%%%%%%%%%%%%%%%%%%%%%%%%%%%%%%%%%%%%%%%%%%%%%%%%%%%%%
We perform simulations using the massively parallel electromagnetic PIC code \TRISTAN \citep{spitkovsky05}. 
We consider Cartesian geometry, including all three components of the particle velocities and of the electromagnetic fields. 
The parameters used in our simulations are listed in Table \ref{tab:simpara} and outlined below.
%%%%%%%%%%%%%%%%%%%%%%%%%%%%%%%%%%%%%%%%%%%%%%%%%%%%%%%%%%%
\subsection{Simulation box and magnetic field}
%%%%%%%%%%%%%%%%%%%%%%%%%%%%%%%%%%%%%%%%%%%%%%%%%%%%%%%%%%%
Most of the simulations are performed in a quasi-1D geometry, with five cells along $y$ and $N_{\rm x}$ cells along $x$; 
the physical length of the box is chosen to be at least $\approx 6\,\lambda_{\rm fast}$ to ensure that the domain spans several wavelenghts of the fastest-growing mode. 
We use $5$ cells per $d_{\rm e}$ and a time step is set by the speed of light and grid space, such that $\Delta t = 0.04\, \omega_{\rm pe}^{-1}$;
we checked the convergence of our results with such resolutions.

Simulations are initialized with a uniform magnetic field in the $x$ direction, whose strength is parameterized via the magnetization $\sigma\equiv (\omega_{\rm ce}/\omega_{\rm p})^2$, where $\omega_{\rm p}=\omega_{\rm pe}(1+m_{\rm e}/m_{\rm i})^{1/2}$;
for our benchmark runs we set $\sigma = 10^{-2}$, i.e., an Alfv\'en speed of $v_{\rm A0}/c\,=(\sigma m_{\rm e}/m_{\rm i})^{1/2}=10^{-2}$.

Although $B_{\rm y,z}=0$ at $t=0$, the thermal motion of the plasma electrons and ions develops a non-zero $B_{\rm \perp}$ after a few time steps, which acts as a seed field for the instability. 
The seed field can be reduced by initializing a smaller plasma temperature at $t=0$; however, we have checked that the final result is unaffected by this choice for relatively cold plasmas \citep[see,e.g.,][for warm-plasma corrections]{reville+08a,zweibel+10}. 

%%%%%%%%%%%%%%%%%%%%%%%%%%%%%%%%%%%%%%%%%%%%%%%%%%%%%%%%%%%
\subsection{Background plasma}\label{subsec:bgplasma}
%%%%%%%%%%%%%%%%%%%%%%%%%%%%%%%%%%%%%%%%%%%%%%%%%%%%%%%%%%%
Each computational cell is initialized with $50$ macro-particles, half representing ions and half electrons. 
An artificial ion to electron mass ratio, $m_{\rm i}/m_{\rm e}=100$, is used to keep the simulations computationally tractable.
Both ion and electron distributions are initialized as Maxwellians with temperature
$T_{\rm i}=T_{\rm e} = 6.4\times 10^{-3} m_ec^2/k_B$, where $k_B$ is the Boltzmann constant.
%this corresponds to a Debye length of $\lambda_{\rm d} = 8\times 10^{-2}\,d_{\rm e}$, where $d_{\rm e}=c/\omega_{\rm pe}$ is the electron skin-depth.
%%%%%%%%%%%%%%%%%%%%%%%%%%%%%%%%%%%%%%%%%%%%%%%%%%%%%%%%%%%
\subsection{Cosmic rays}
%%%%%%%%%%%%%%%%%%%%%%%%%%%%%%%%%%%%%%%%%%%%%%%%%%%%%%%%%%%
To be in the NRSI regime, $n_{\rm cr} \ll n_{\rm 0}$ is needed so, to boost the CR counting statistics, we use $25$ CR particles per cell with weights tuned to set the ratio $n_{\rm cr}/n_{\rm 0}$ as described in Table \ref{tab:simpara} \citep[see, e.g.][]{riquelme+09};
to retain the quasi-neutrality, the weights of the background electrons are either increased or decreased depending on the sign of the CRs.
%\SG{If $\alpha_{\rm i}$, $\alpha_{\rm e}$, and $\alpha_{\rm cr}$ represent the weight of ions, electrons, and CRs respectively then for CR-ion case, we use $\alpha_{\rm e}=\alpha_{\rm i}+\alpha_{\rm cr}$ and for CR-electron case $\alpha_{\rm e}=\alpha_{\rm i} - \alpha_{\rm cr}$, where the default value of $\alpha_{\rm i}$ and $\alpha_{\rm e}$ is unity. In our fiducial runs, $\alpha_{\rm cr}=4\times 10^{-3}$, which mimics $n_{\rm cr}/n_{\rm 0}=4\times 10^{-3}$.}
This means that in the CR-ion (CR-electron) case, the thermal plasma contains a slightly larger number of electrons (ions).
For all three species (ion, electrons, and CRs), the initial spatial distribution of macro-particles in a computational cell is the same, which ensures a zero electric field at $t=0$.

In the reference frame in which CR are isotropic, they have momentum  $p^{\prime}_{\rm cr}=\gamma_{\rm cr}m_{\rm cr} v^{\prime}_{\rm cr}$ (where $v^{\prime}_{\rm cr}$ is the isotopic velocity);  for a meaningful comparison between the CR-I and CR-E cases, we use the same CR momentum for both species (see Equation \ref{eq:nrlimit}).

The isotropic CR distribution is boosted with velocity $v_{\rm bst}$ with respect to the background thermal ions, which are initially at rest;
thermal electrons have a drift velocity defined by Equation \ref{eq:veloe}.
Due to the Lorentz transformation, the effective drift velocity between CRs and thermal plasma along the $x$ axis becomes:
\begin{equation}\label{eq:vdeff}
v_{\rm d} =\frac{1}{2}\int_{\rm -1}^{1} d\mu \frac{\mu\,v^{\rm \prime}_{\rm cr}+v_{\rm bst}}{1+\mu\,v^{\rm \prime}_{\rm cr}\,v_{\rm bst}/c^2}
\end{equation}
Note that the boost velocity is not identical to the drift velocity. In the simulation frame, the average momentum per particle is also modified to 
\begin{equation}\label{eq:pcrlab}
p_{\rm cr, x} = \frac{1}{2}\int_{\rm -1}^{1} d\mu \frac{\mu\, p^{\prime}_{\rm cr}+E^{\prime}_{\rm cr}\,v_{\rm bst}/c^2}{[1-(v_{\rm bst}/c)^2]^{1/2}} = \gamma_{\rm bst}\,v_{\rm bst} \frac{E^{\prime}_{\rm cr}}{c^2},
\end{equation}
where $E^{\prime}_{\rm cr}=\gamma_{\rm cr}m_{\rm cr} c^2$, $\gamma_{\rm bst}=1/[1-(v_{\rm bst}/c)^2]^{1/2}$. For our fiducial parameters: $v_{\rm bst}=0.8\,c$, $|p^{\prime}_{\rm cr}|= 10\,m_{\rm i}\,c$, we find $v_{\rm d}\simeq 0.635\,c$ and $p_{\rm cr,x}\simeq13.4\,m_{\rm i}\,c$, which yield $\xi\approx 340$ (Equation \ref{eq:nrlimit}).
For $t>0$, all species are allowed to evolve self consistently under periodic boundary conditions.
%%%%%%%%%%%%%%%%%%%%%%%%%%%%%%%%%%%%%%%%%%%%%%%%%%%%%%%%%%%
%%%%%%%%%%%%%%%%%%%%%%%%%%%%%%%%%%%%%%%%%%%%%%%%%%%%%%%%%%%
%%%%%%%%%%%%%%%%%%%%%%%  Section 4  %%%%%%%%%%%%%%%%%%%%%%%
%%%%%%%%%%%%%%%%%%%%%%%%%%%%%%%%%%%%%%%%%%%%%%%%%%%%%%%%%%%
%%%%%%%%%%%%%%%%%%%%%%%%%%%%%%%%%%%%%%%%%%%%%%%%%%%%%%%%%%%
\section{Results}\label{sec:results}
%%%%%%%%%%%%%%%%%%%%%%%%%%%%%%%%%%%%%%%%%%%%%%%%%%%%%%%%%%%
%%%%%%%%%%%%%%%%%%%%%%%%%%%%%%%%%%%%%%%%%%%%%%%%%%%%%%%%%%%
\subsection{Maximally-unstable modes}\label{subsec:maximally}
\begin{figure}
\centering
\includegraphics[height=2.0in,width=3.5in]{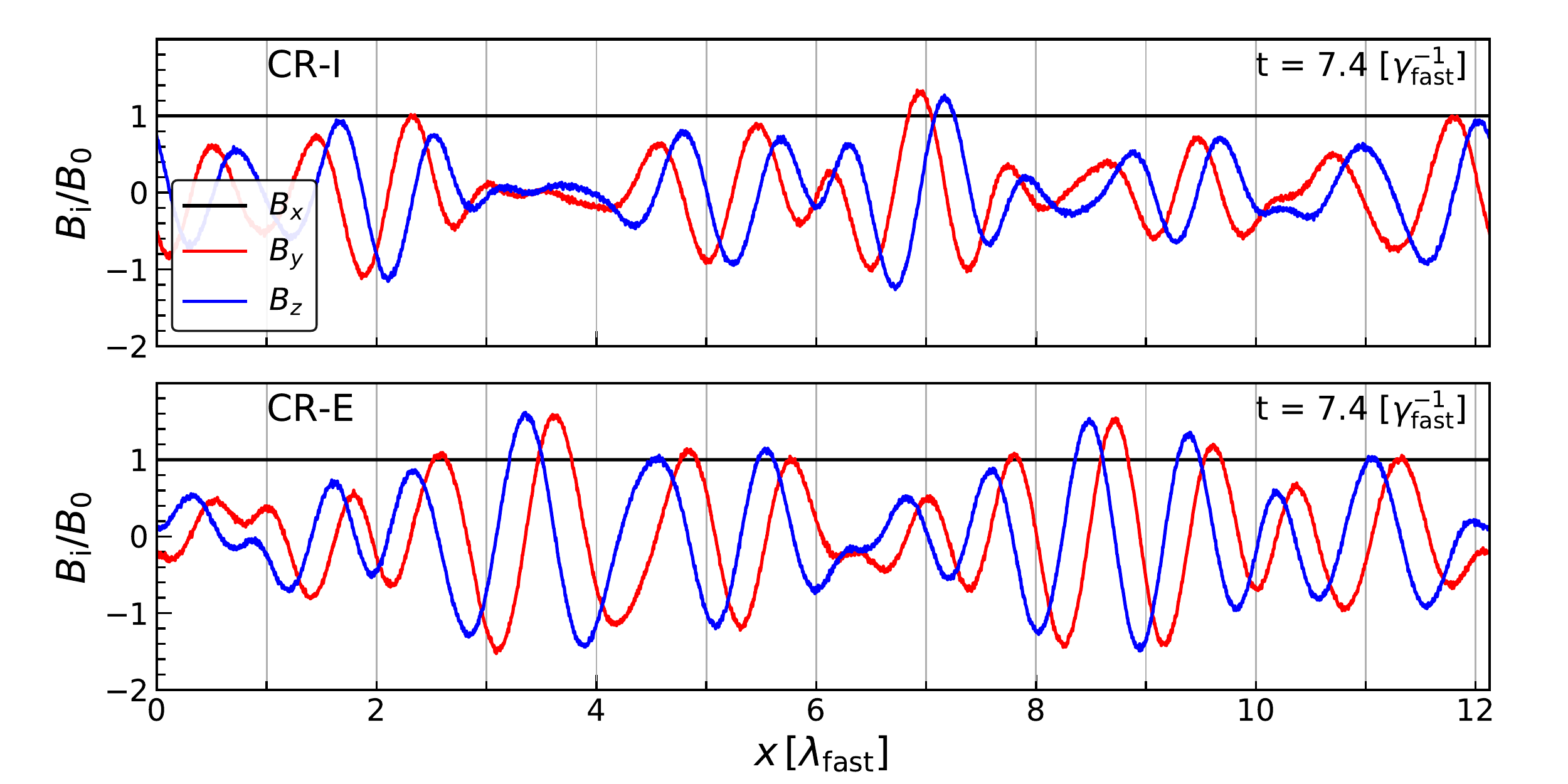}
\caption{Snapshot of magnetic field component at $t\simeq7.4\, \gamma_{\rm fast}^{-1}$ as a function of $x$, in units of  $\lambda_{\rm fast}\simeq 494.7\,d_{\rm e}$, for the benchmark run A. EI-S-$\xi340$ (Table \ref{tab:simpara}).
Top and bottom panels display CR-ion (CR-I) and CR-electron (CR-E) cases, respectively. For both cases, a typical mode of wavelength $\approx\lambda_{\rm fast}$ is evident.} \label{fig:profcomp}
\end{figure}

%%%%%%%%%%%%%%%%%%%%%%%%%%%%%%%%%%%%%%%%%%%%%%%%%%%%%%%%%%%
The magnetic field profiles for our benchmark parameters (Run A in Table \ref{tab:simpara}) are shown in Figure \ref{fig:profcomp}, which are taken at $t\simeq 7.4\,{\rm \gamma_{\rm fast}^{-1}}$;
where both times and lengths are normalized to the prediction for the fastest growing mode ($\gamma_{\rm fast}^{-1}$ and $\lambda_{\rm fast}\equiv 2\pi/k_{\rm fast}$, see Equations \ref{eq:kmax} and \ref{eq:wmax}). 
Black, red, and blue lines correspond to the $x,y$, and $z$ components of ${\bf B}$ normalized to $B_{\rm 0}$.
While $B_{\rm x}$ cannot change in a quasi-1D setup, the perpendicular components show a dominant mode with a wavelength of $\approx \lambda_{\rm fast}$ consistent with Equation \ref{eq:kmax}. 
Comparing the CR-I and CR-E driven runs (top and bottom panels of Figure \ref{fig:profcomp}, respectively), we see that magnitude and wavelength of the dominate mode are very similar.
However, in the top panel, $B_{\rm z}$ (blue) leads $B_{\rm y}$ (red), while in the bottom panel it trails  $B_{\rm y}$;
 this is associated with the helicity of the growing modes, consistent with Equation \ref{eq:disp_0}.
 
%##########################################
%##########################################
%%%%%%%%%%%%%%%%%%%%%%%%%%%%%%%%
%\subsection{Phase difference and helicity}
%%%%%%%%%%%%%%%%%%%%%%%%%%%%%%%%
The helicity of each mode with wavenumber $k$ can be formally expressed by the phase difference of the perpendicular magnetic fields, $\Delta\phi(k)$ (Equation \ref{eq:disp_0}), written as a function of Stoke's parameters ($Q,U,V$; see Equation \ref{eq:appstokes} in Appendix \ref{app:disper}): 
\begin{equation}\label{eq:polangle}
\Delta\phi(k) = -\tan^{-1}\left[\frac{V(k)}{(Q^2(k)+U^2(k))^{1/2}}\right] \ ,
\end{equation}
For a given $k$, the helicity depends on the sign of $\Delta\phi$; a mode is RH if $\Delta\phi(k) > 0$ and LH if $\Delta\phi < 0$ and the polarization is exactly circular if $|\Delta\phi|=\pi/2$.

\begin{figure}
\centering
\includegraphics[width=0.47\textwidth]{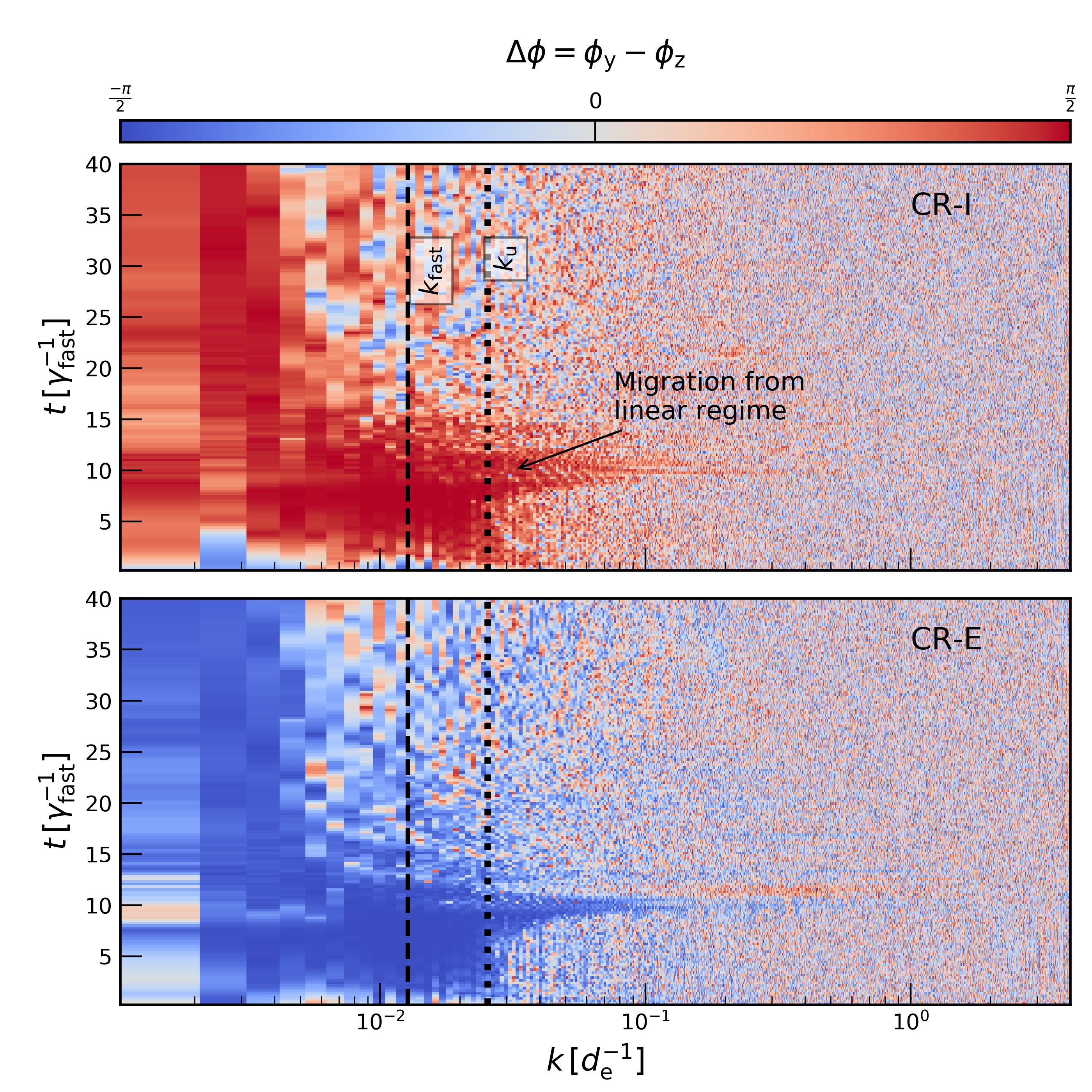}
\caption{
Time evolution of the phase difference $\Delta\phi$ (a proxy for helicity, see Equation \ref{eq:polangle}) for our benchmark run. 
Top and bottom panels display CR-I and CR-E cases. 
The two vertical lines, dashed and dotted, denote $k_{\rm fast}=1.27\times 10^{-2}\,d_{\rm e}^{-1}$ and $k_{\rm u}=2k_{\rm fast}$. 
For $k>k_{\rm u}$ (regime I in Figure \ref{fig:diffregime}) waves do not have a fixed helicity, whereas for $k<k_{\rm u}$ (regime II), $\Delta\phi$ is either $\approx +\pi/2$ or $-\pi/2$, corresponding to RH (CR-I case) or LH (CR-E case) modes. 
Deviation from the linear theory is observed for $t \gtrsim 10\, \gamma_{\rm fast}^{-1}$.
}
\label{fig:polangle}
\end{figure}

The phase difference $\Delta\phi(k)$ is shown in Figure~\ref{fig:polangle} as a function of time for CR-I (upper panel) and CR-E (lower panel) cases.
For $k\lesssim k_{\rm u}$ (left of the vertical dotted line), we have that $\Delta\phi\to \pm \pi/2$ for CR-I and CR-E cases, consistent with expectations of RH and LH modes, respectively.
For $k>k_{\rm u}$, modes do not have a fixed mode of polarization, in that both branches have a comparable amplitude and do not grow in the linear stage compared to other modes (cf. \S \ref{subsec:result-growth}). 

\begin{figure}
\centering
\includegraphics[height=3.7in,width=3.35in]{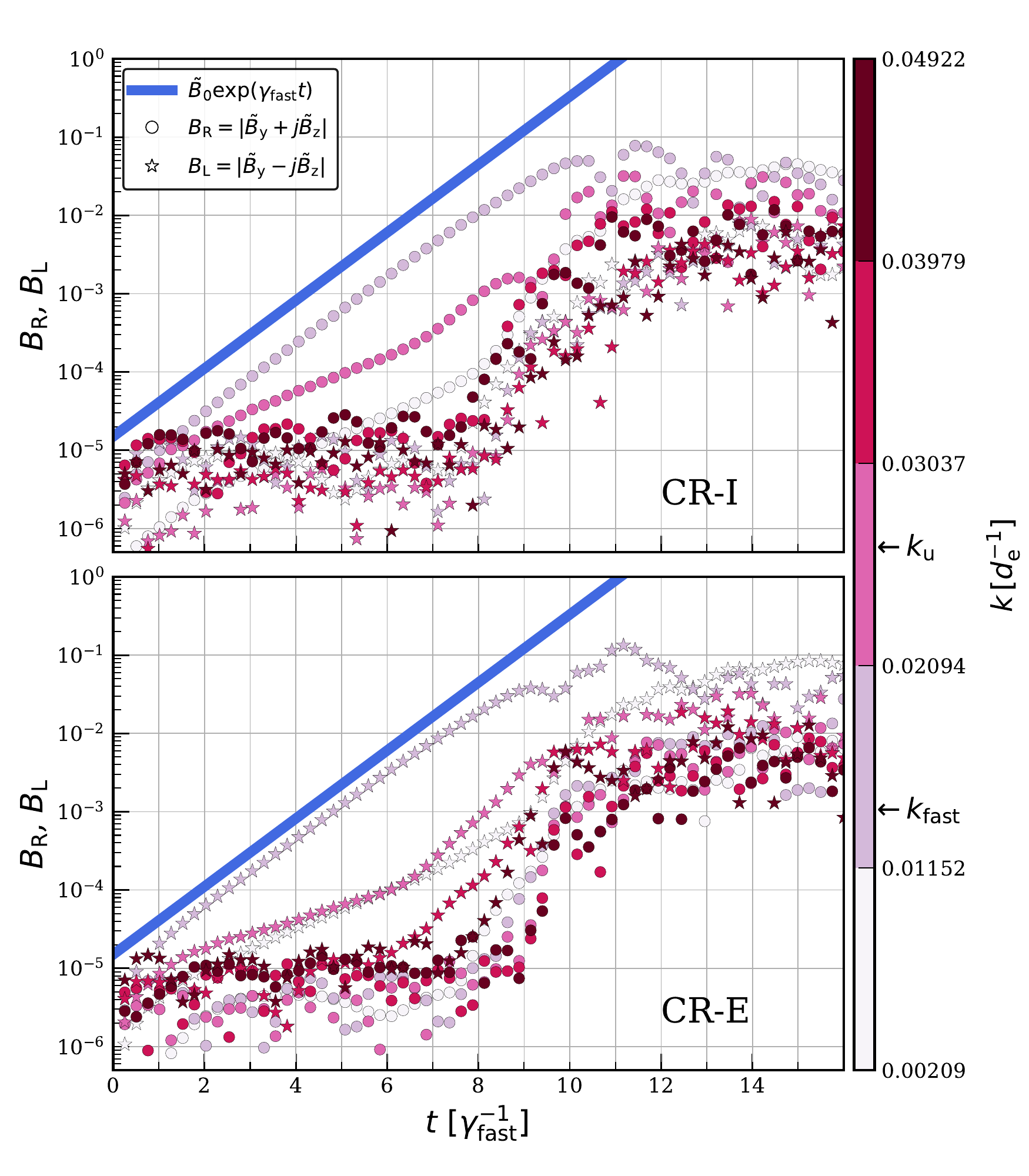}
\caption{Time evolution of LH ($B_L$, stars) and RH ($B_R$, circles) modes for five different wavenumbers $k$. 
In the CR-I case, RH modes grow the fastest, whereas in the CR-E case the magnetic field is dominated by growing LH modes.
In both cases, the value of $k$ for the fastest growing mode is $\approx 0.0127\, d^{-1}_{\rm e}$ (lilac). 
The simulation shows a good agreement with the analytic prediction for $k=k_{\rm fast}$ (blue solid line) until $t\approx 9\,{\rm \gamma_{\rm fast}^{-1}}$, after which the evolution becomes non-linear. Note that when $t\gtrsim 12\,{\rm \gamma_{\rm fast}^{-1}}$, the amplitude of the fastest growing mode becomes sub-dominant and longer wavelength modes take over.}\label{fig:Bk}
\end{figure}

By looking at the time evolution of $\Delta\phi$ (vertical axis in Figure \ref{fig:polangle}), we find that after $t\approx 9{\rm \gamma^{-1}_{\rm fast}}$, the red/blue regions deviate from the linear prediction (vertical dashed/dotted line). 
This is due to the CR back-reaction:
the thermal plasma is set in motion (see e.g., Equation \ref{eq:pacc} in Appendix \ref{app:sat}) and $v_{\rm d}$ is reduced, modifies the upper limit, $k_{\rm u}$. 
At $t\approx 9{\rm \gamma^{-1}_{\rm fast}}$ the resonant branch also starts to grow very rapidly (Figure \ref{fig:Bk}) and the helicity is no longer sharp.
When $t\gtrsim 12{\rm \gamma^{-1}_{\rm fast}}$, the system becomes non-linear.

%##########################################
%##########################################
%##########################################
%##########################################
%%%%%%%%%%%%%%%%%%%%%%%%%%%%%%%%
\subsection{Growth rate} \label{subsec:result-growth}
%%%%%%%%%%%%%%%%%%%%%%%%%%%%%%%%
To compare the growth rate in the CR-I and CR-E cases, in Figure \ref{fig:Bk} we show the time evolution of RH and LH modes ($B_{\rm R}\equiv \tilde{B}_{\rm y}+j\tilde{B}_{\rm z}$, circles, and $B_{\rm L}\equiv\tilde{B}_{\rm y}-j\tilde{B}_{\rm z}$, stars; where $\tilde{B}_{\rm y,z}(k)$ are the Fourier transform of $B_{\rm y,z}(x)$ along the $x$ axis) for different values of $k$. 
Again, we see that in the CR-I  case (top panel) RH modes with $k\lesssim k_{\rm u} = 0.0254\,d_{\rm e}^{-1}$ grow faster than their RL counterparts until $t \approx 9\,\gamma_{\rm fast}^{-1}$; 
the opposite is true for the CR-E case (bottom panel).
A comparison between the blue solid line (showing the expected evolution of the fastest growing mode) and purple coloured circles (upper panel) or stars (lower panel) for $k_{\rm fast}=0.0127\,d_{\rm e}^{-1}$ indicates that the growth rate of the fastest-growing mode is the same for both cases, consistent with Equation \ref{eq:wmax}. 
Note that as long as modes remain quasi-linear ($t \lesssim 9\,\gamma_{\rm fast}^{-1}$, modes with $k>k_{\rm u}$ (red/brown circles/stars) in the non-unstable branch just oscillate, as suggested in section \ref{subsec:nonreso-limit}.
For $t \gtrsim 9\,\gamma_{\rm fast}^{-1}$, both RH and LH modes evolve similarly, likely because of power transfer between modes of different helicities \citep[e.g.,][]{chin-wentzel72}, when the system has entered its non-linear regime (also see Figure \ref{fig:polangle}).
%%%%%%%%%%%%%%%%%%%%%%%%%%%%%%%%%%%%%%%%%%%%%%%%%%%%%%%%%%%%%%%%
\begin{figure*}
\centering
\includegraphics[width=\textwidth]{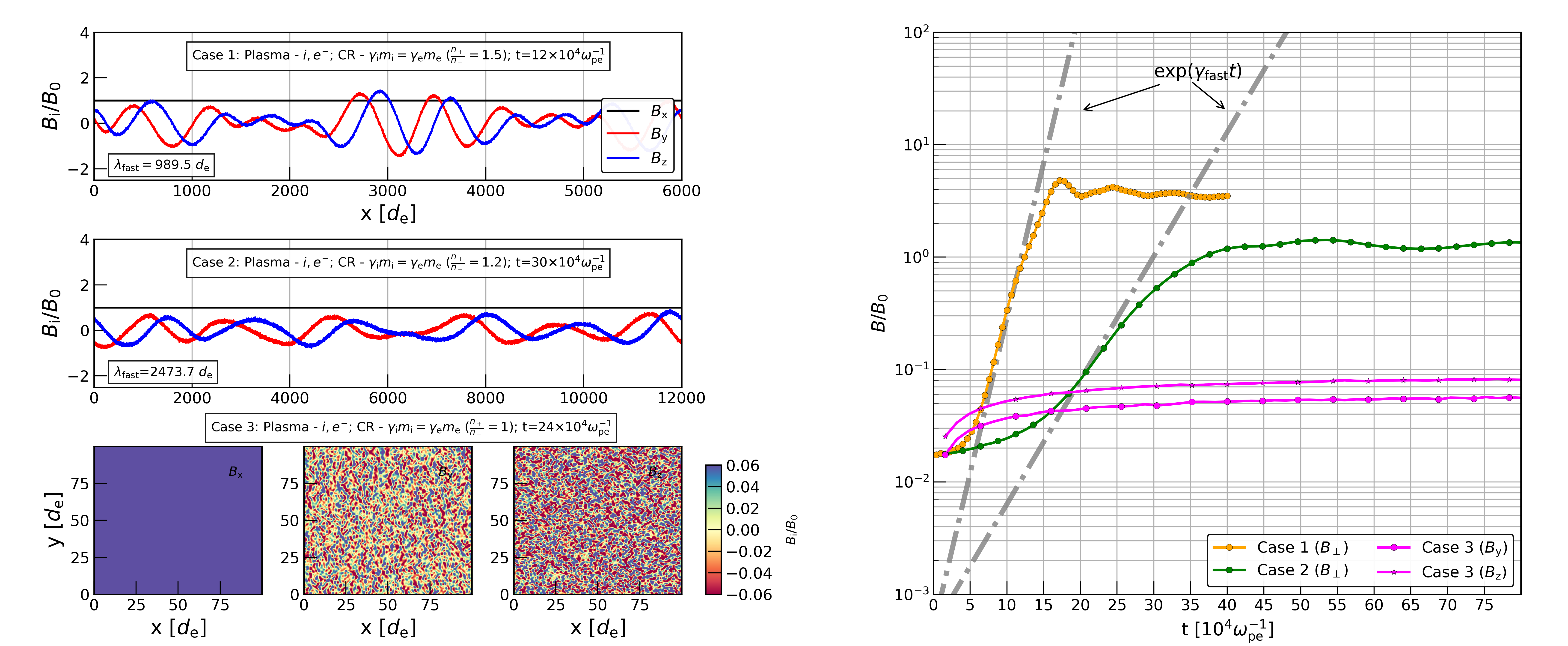}
\caption{NRSI driven by pair beams in a ion-electron background (Runs E-G in Table \ref{tab:simpara}). 
Left panels: snapshots of $B$-field for three cases with different ratio $n_{\rm +}/n_{\rm -}$. 
Right panel: time evolution of the $B$-field for corresponding cases; 
the exponential phase is well described by the linear theory outlined here, except for the zero-current case $3$ (Run G).}\label{fig:NRSI-eibck}
\end{figure*}
%%%%%%%%%%%%%%%%%%%%%%%%%%%%%%%%%%%%%%%%%%%%%%%%%%%%%%%%%%%%%%%%

In summary, the electron-driven NRSI produces result similar to ion-driven case when $\gamma_{\rm e}$ in the CR beam $\approx m_{\rm i}/m_{\rm e} \gamma_{\rm i}$. Next we use this result to explore NRSI in other environments where the NRSI can be potentially important.
%%%%%%%%%%%%%%%%%%%%%%%%%%%%%%%%%%%%%%%%%%%%%%%%%%%%%%%%%%%%%%%%
\subsection{NRSI in different environments}\label{subsec:diffback}
%%%%%%%%%%%%%%%%%%%%%%%%%%%%%%%%%%%%%%%%%%%%%%%%%%%%%%%%%%%%%%%%
In previous sections, we have presented the cases where the CR populations are comprised entirely of either ions or electrons. 
However, in some astrophysical environments, energetic particles consist of both energetic positrons and electrons and the thermal background can be a pair plasma.
If there is a difference in acceleration efficiency between these two species \citep[e.g.,][]{cerutti+15,philippov+18}, then they can generate a current, which may drive the NRSI. 
When such relativistic electrons are liberated into the interstellar medium (an electron--ion plasma), they may excite the NRSI and amplify magnetic field that may be crucial for the self-confinement of CRs near their sources, as revealed, e.g., by the $\gamma$-ray halos detected around PWNe (e.g., \citealt{hawc17}).

Denoting the number density of positive and negative charges by $n_{\rm +}$ and $n_{\rm -}$ respectively, the linear theory predicts that the growth of the NRSI depends on the effective CR current density, $J_{\rm cr}\equiv (n_{\rm +}-n_{\rm -}) e\,v_{\rm d}$, which physically corresponds to the return current in the background plasma.
However, since the helicity of waves excited by positrons and electrons are opposite, PIC simulations are necessary to assess the extent to which a pair beam can be viewed as a linear superposition of their opposite currents. 
To cover different scenarios, we now investigate the NRSI driven by CRs of both charges on top of two different thermal backgrounds: ion-electron (\S \ref{subsubsec:ieplasma}) and electron-positron plasmas (\S \ref{subsubsec:pairplasma}).

%%%%%%%%%%%%%%%%%%%%%%%%%%%%%%%%%%%%%%%%%%%%%%%%%%%%%%%%%%%%%%%%
\begin{figure*}
\centering
\includegraphics[width=\textwidth]{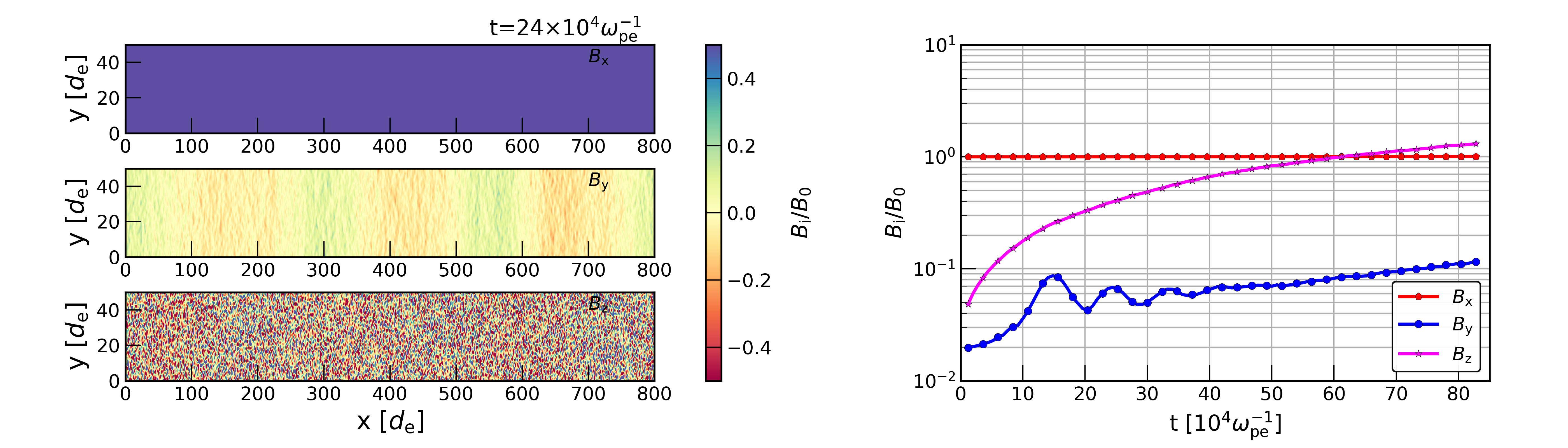}
\caption{NRSI in pair plasma for a zero beam-current case (Run I in Table \ref{tab:simpara}).
Left panels: components of B at $t=24\times 10^4\,\omega_{\rm pe}^{-1}$, showing filaments of size $\sim d_{\rm e}$. 
Right panel: time evolution of corresponding components;
the growth of the instability is very different from the standard theory of NRSI.
}\label{fig:NRSI-epbck}
\end{figure*}
%%%%%%%%%%%%%%%%%%%%%%%%%%%%%%%%%%%%%%%%%%%%%%%%%%%%%%%%%%%%%%%%
%%%%%%%%%%%%%%%%%%%%%%%%%%%%%%%%%%%%%%%%%%%%%%%%%%%%%%%%%%%%%%%%
\subsubsection{Pair beam in an ion-electron plasma}\label{subsubsec:ieplasma}
%
%%%%%%%%%%%%%%%%%%%%%%%%%%%%%%%%%%%%%%%%%%%%%%%%%%%%%%%%%%%%%%%%
We first consider an ion-electron background plasma with $m_{\rm i}/m_{\rm e}=100$ (as in previous sections), and CRs with the Lorentz factors $\gamma_{\rm i}m_{\rm i}=\gamma_{\rm e}m_{\rm e}=10$. 
%These choices make CRs analogous to the energetic pair beam and the background to the ion-electron plasma. 
We allow  $n_{\rm +}$ and $n_{\rm -}$ to be different, since positrons and electrons can be accelerated in different regions with different net electric charges.
For instance, in the equatorial region of a pulsar magnetosphere the reconnecting current sheet produces more energetic positrons than electrons for an aligned rotator\footnote{The opposite would be true for an anti-aligned rotator, where the angle between magnetic and rotation axes is $\sim \pi$ instead of 0.} \citep[][]{cerutti+15,philippov+18}. 
Even if the ultimate mechanism responsible for the acceleration of the bulk leptons that shine in a PWNe is still under debate, it is arguable that such magnetospheric particles play a crucial role, likely acting at least as seeds for further acceleration, possibly at the wind termination shock.
Therefore, ``pair'' beams in and around PWNe may be either neutral or present an excess of particles of one sign.

Let us first consider the regime $n_{\rm +}> n_{\rm -}$, and more precisely two cases in which there are $50\%$ and $20\%$ more positively-charged particles (labelled by case 1 and case 2 in Figure \ref{fig:NRSI-eibck}, respectively; the corresponding parameters are detailed in runs E and F of Table \ref{tab:simpara}).
The snapshots of the B-field for these two cases are shown in the top- and middle-left panels of Figure \ref{fig:NRSI-eibck}.
We find that the wavelength and growth rate of the fastest growing mode agree well with the linear theory when an effective number density of CRs $n_{\rm cr}=n_{\rm +}-n_{\rm -}$ is used.
This is shown by the grey dash-dotted and dotted lines in the right panel in the same Figure, which displays the evolution of B  in time for both cases. 
Note that for lower effective currents ($20\%$ excess, green curve) the growth rate is smaller and also the saturation of the NRSI occurs at smaller values, still $B_{\rm \perp}/B_{\rm 0}\gtrsim 1$ for our parameters.

The third case considers the scenario $n_{\rm +}= n_{\rm -}$, where we observe that the NRSI is quenched, as expected from the linear theory for a null CR current.
This can be seen in the lower-left panels of Figure \ref{fig:NRSI-eibck} (Run G in Table \ref{tab:simpara}) and also from the right panel of the same figure (magenta curves). 
Note that the system still has free energy because of the CR anisotropy, and in fact we observe evidence of small-scale fluctuations and a marginal amplification of the magnetic field, possibly associated with the gyro-resonant instability discussed by \citet{lebiga+18}.

This situation may be more similar to the case of the relativistic beams of pairs produced by the interaction of blazar TeV photons with the extragalactic background light, though in a significantly more magnetized background plasma \citep[the electrostatic oblique instability, see e.g.,][and references therein]{sironi+14,shalaby+17}.
A more detailed investigation of this regime is left to a further work, but here we stress that even a relatively small excess of one charge with respect to the other, as naturally expected from pulsars, is likely sufficient to put the system in the Bell (or resonant) regime.
%%%%%%%%%%%%%%%%%%%%%%%%%%%%%%%%%%%%%%%%%%%%%%%%%%%%%%%%%%%%%%%%
\subsubsection{Beams in pair plasmas}\label{subsubsec:pairplasma}
%
%%%%%%%%%%%%%%%%%%%%%%%%%%%%%%%%%%%%%%%%%%%%%%%%%%%%%%%%%%%%%%%%
Let us now consider the development of the NRSI in a pair plasma (runs H and I in Table \ref{tab:simpara}).
At first we investigate the effect of a background pair plasma on the standard NRSI; 
we take the current to be made of only positively-charged particles, i.e., positrons, and therefore expect results similar to the ion-driven cases.
While estimating the growth rate, one has to recall that posing $m_{\rm i}=m_{\rm e}$ reduces $k_{\rm fast}$ and $\gamma_{\rm fast}$ by a factor of $2^{1/2}$ and $2$ compared to the standard ($m_{\rm i}\gg m_{\rm e}$) prediction (Equations \ref{eq:kmax} and \ref{eq:wmax} respectively). 
These factors are due to the fact that $v_{\rm A0}$ in Equation \ref{eq:kmax} is practically $v_{\rm A0i}=v_{\rm A0}(1+m_{\rm e}/m_{\rm i})^{1/2}$, and $\mathcal{A}$ in Equation \ref{eqapp:disp_full2} is 2 instead of 1 (for details, see Appendix \ref{app:disper}).
The simulations that we performed in this regime confirm such theoretical estimates and easily produce $B_{\rm \perp}/B_{\rm 0}\gtrsim 1$ as expected, so we do not show them here.

For a pair background, it is possible to envision a scenario \citep[e.g., in relativistic shocks, see][]{sironi+09}, in which both electrons and positrons are accelerated in the same way and the effective current in CRs is zero. 
This case (Run I in Table \ref{tab:simpara}) is illustrated in Figure \ref{fig:NRSI-epbck}, which displays the three components of the magnetic field (left panels) and their time evolution (right panel).
We point out that there are substantial differences between the electron-ion (Figure \ref{fig:NRSI-eibck}) and pair (Figure \ref{fig:NRSI-epbck}) backgrounds.
Unlike in the electron-ion case, the out-of-plane component $B_z$ does not saturate at $\delta B\ll B_0$ but grows over the whole simulation;
$B_y$ grows with a similar rate, too, but it smaller by a factor of a few, likely as a consequence of the reduced dimensionality of the simulation.
This is consistent with the PIC simulations of relativistic shocks in pair plasmas performed by \citealt{sironi+09}, where electrons and positrons are equally accelerated and produce non-linear fluctuations in the shock precursor. 
We also note that, while fluctuations in $B_z$ have very small wavelengths, of the order of the inertial length in both the longitudinal and transverse direction (similar to the case in Figure \ref{fig:NRSI-eibck}), there is a clear evidence of a long-wavelength longitudinal mode in $B_y$.

The possibility of developing large-scale (i.e., much larger than $d_e$) non-linear fluctuations even for a case with zero-current is indeed intriguing and may have astrophysical implications for the self-confinement of energetic pairs.
In any case, this instability is quite different from the NRSI in many aspects, and the anisotropy that we report is likely an artifact of the reduced dimensionality of the presented simulations.
A dedicated investigation of this regime with 3D runs is in order but beyond the goals of this paper. 
%%%%%%%%%%%%%%%%%%%%%%%%%%%%%%%%%%%%%%%%%%%%%%%%%%%%%%%%%%%%%%%%
\subsection{Saturation} \label{subsec:result-saturation}
%%%%%%%%%%%%%%%%%%%%%%%%%%%%%%%%%%%%%%%%%%%%%%%%%%%%%%%%%%%%%%%%
The NRSI is believed to be important for the overall amplification of an initial magnetic field, and the exact mechanism for its saturation is not completely understood. 
\citet{bell04} and \citet{blasi+15} provided two different heuristic arguments for deriving the expected strength of the amplified magnetic field, which converge in suggesting that at saturation
\begin{equation}\label{eq:bellsat}
    \frac{\delta B^2}{4\pi}\approx \frac{v_{\rm d}}{c}\,U_{\rm cr}
\end{equation}
This condition\footnote{For a shock,  $\xi\approx \epsilon \mathcal{M}_{\rm A}^2 (v_{\rm d}/c)$, where $\mathcal{M}_{\rm A}$ is the Alfv\'{e}n Mach number, $v_{\rm d}$ is the speed of the shock in upstream frame, and $\epsilon=U_{\rm cr}/(\rho v_{\rm d}^2) \sim 0.1$ is CR acceleration efficiency \citep[e.g.,][]{caprioli+14b}.} is similar to posing $\xi\approx 1$ in Equation \ref{eq:nrlimit}, since $P_{\rm cr}\approx v_{\rm d} U_{\rm cr}/c$, which is also equivalent to stating that when the RSI and the NRSI grow at the same rate, the CR current is disrupted and perturbations cannot grow linear.
On the other hand, kinetic simulations \citep[e.g.,][]{riquelme+09,gargate+10,caprioli+14b,Weidl+19a} suggested that saturation may be achieved when modes that can  scatter the CRs have grown sufficiently, a statement that is hard to quantify in the non-linear stage; 
therefore, the question arises whether CR-I and CR-E NRSI evolve and saturate in a similar way.

%%%%%%%%%%%%%%%%%%%%%%%%%%%%%%%%%%%%%%%%%%%%%%%%%%%%%%%%%%%%
\begin{figure*}
\centering
\includegraphics[width=0.99\textwidth]{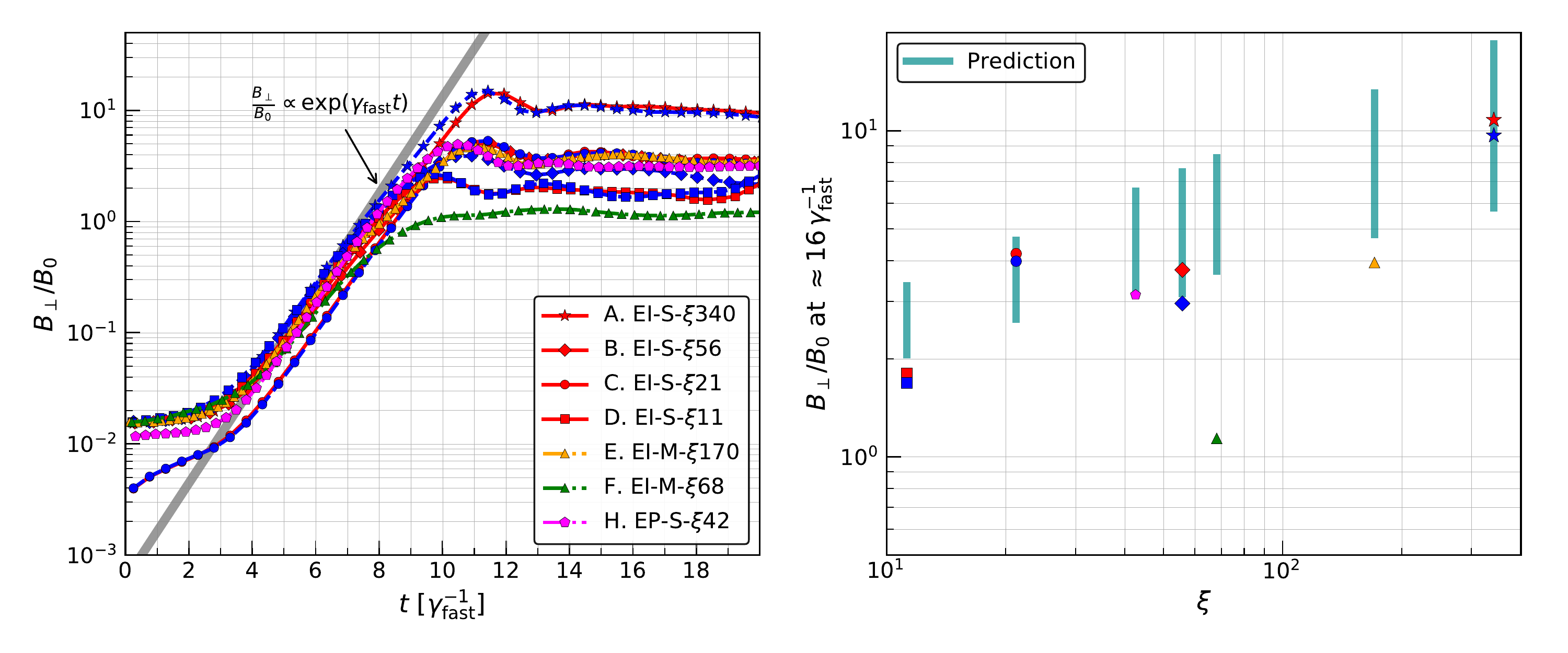}
\caption{
Left panel: time evolution of the box-averaged transverse field $B_{\rm \perp}/B_{\rm 0}$ for all runs (except the zero beam current runs G and I) listed in Table \ref{tab:simpara}. 
The grey line shows the expected linear growth. 
Comparing the grey lines with our simulations, we find that for CR-E (blue dashed curves) and CR-I (red solid curves) cases, $B$ evolves similarly and saturates at the same level. 
Right panel:  $B_{\rm \perp}/B_{\rm 0}$ at saturation ($t\sim 16\,\gamma_{\rm fast}^{-1}$) as a function of the $\xi$ parameter, a proxy for NRSI prominence over RSI (see Equation \ref{eq:nrlimit} and Table \ref{tab:simpara}). 
The cyan lines show the expectations from Appendix \ref{app:sat}.
}\label{fig:allin1}
\end{figure*}
%%%%%%%%%%%%%%%%%%%%%%%%%%%%%%%%%%%%%%%%%%%%%%%%%%%%%%%%%%%%
To investigate the saturation of the magnetic field, we explore different plasma and CR parameters such as $v_{\rm A0}$, $n_{\rm cr}/n_{\rm e}$, $v_{\rm d}$, and $p_{\rm cr}$ (see Table \ref{tab:simpara}) and display the evolution of the transverse  magnetic field in Figure \ref{fig:allin1}.
All the simulations have $\xi\gg 1$ and in fact are conducive to $B_{\rm \perp}/B > 1$. 
By comparing the red and blue curves (representing CR-I and CR-E cases), we conclude that the time evolution and the saturation of magnetic field amplification depends only on the dynamic mass of CR particles, and not on the charge of the CR current. 

For a qualitative estimate of the saturated magnetic field in our simulations, we extend our linear analysis by using a semi-classical approach (Appendix \ref{app:sat}), which is compared with simulations and displayed in the right panel of Figure \ref{fig:allin1}. 
The top of the cyan lines represents the upper limit of the final $B_{\rm \perp}/B_{\rm 0}$, which matches Equation \ref{eq:bellsat}. 
Note that saturation may be slightly different if CRs were continuously replenished, rather than obeying periodic boundary conditions as in the present setup.
Although Figure \ref{fig:allin1} shows a reasonable agreement with theoretical prediction, we want to draw attention to the cases where the mixed composition of CRs are shown (in particular Run F -- green triangle). 
The saturated $B_{\rm \perp}/B_{\rm 0}$ for these runs is appreciably smaller than the prediction, as mentioned above.
An important result is that the NRSI, whether driven by a mixed CR composition or in a different background plasma, and typically results in $B_{\rm \perp}/B_{\rm 0}\gtrsim 1$.
%%%%%%%%%%%%%%%%%%%%%%%%%%%%%%%%%%%%%%%%%%%%%%%%%%%%%%%%%%%
%%%%%%%%%%%%%%%%%%%%%%%%%%%%%%%%%%%%%%%%%%%%%%%%%%%%%%%%%%%
%%%%%%%%%%%%%%%%%%%%%%%  Section 6  %%%%%%%%%%%%%%%%%%%%%%%%%%%%%%
%%%%%%%%%%%%%%%%%%%%%%%%%%%%%%%%%%%%%%%%%%%%%%%%%%%%%%%%%%%
%%%%%%%%%%%%%%%%%%%%%%%%%%%%%%%%%%%%%%%%%%%%%%%%%%%%%%%%%%%
\section{Summary}\label{sec:summary}
We have investigated the non-resonant streaming instability (NRSI) for different charge, mass, and mixed compositions of CRs in different backgrounds. 
We performed a linear analysis in \S\ref{sec:analytics} and confirmed the analytic predictions using self-consistent PIC simulations. 
Our results are summarized in the following.

\begin{compactitem}
\setlength\itemsep{0.1em}
\item Regardless of the nature of the current-carrying species, the main requirement for driving NRSI, and hence non-linear field amplification, is that the CR momentum flux must be much larger than the magnetic pressure in the background plasma (Equation \ref{eq:nrlimit}). 

\item The growth rate in the CR-I and CR-E cases are comparable at a fixed current, but the helicity of the unstable modes is opposite in the CR ion- and electron-driven cases (Figure \ref{fig:polangle});
this is a consequence of the opposite sign of the return current in thermal electrons that compensates the CR current (Figure \ref{fig:schdia}).

\item  A beam encompassing both positive and negative charges can drive the NRSI and lead to non-linear field amplification, as long as it has a net current, which determines the actual growth rate (Figure \ref{fig:NRSI-eibck}).

\item For a given CR current made of one species only, the magnetic field at saturation ($\delta B/ B_{\rm 0} > 1$) depends on the initial anisotropic momentum flux, and not on its charge (Figure \ref{fig:allin1}). 
This point suggests that laboratory experiments, with sufficiently powerful lasers \citep[e.g.,][]{jao+19}, may be able to test the Bell instability even with electron beams.

\item For CR distributions with the same momentum flux, but encompassing  different charges, less magnetic field is found at saturation (Figure \ref{fig:NRSI-eibck}).
This is a promising path for explaining the origin of the TeV halos detected around PWNe \citep{hawc17}, which are likely produced by escaping energetic leptons. 
The extent of such halos is consistent with a suppression of the Galactic diffusion coefficient of a factor of $\sim 100$, which may be achieved even with linear field amplification, $\delta B/B_0\lesssim 1$.

\item The NRSI driven by a net current behaves in a similar way in ion-electron and in pair plasmas, which is non-trivial due to the different nature of the return current in the background plasma (Figures \ref{fig:NRSI-eibck} and \ref{fig:NRSI-epbck}).
One notable difference is found for the case of a pair beam in a pair plasma, which exhibits more magnetic field amplification than its counterpart in a electron-ion background (Figure \ref{fig:NRSI-epbck}).
\end{compactitem}

In summary, we have provided a theory/simulation cookbook for the properties of the NRSI (Bell) instability for beams and background made of different species, covering a region of the parameter space that ---to our knowledge--- had never been tested via kinetic plasma simulations. 
Applications to given space/astro/laboratory environments will be presented in future works.

\begin{center}
   {\it Software:} \TRISTAN (\citealt{spitkovsky05}).
\end{center}

\section*{ACKNOWLEDGMENTS}
Simulations were performed on computational resources provided by the University of Chicago Research Computing Center, the NASA High-End Computing Program through the NASA Advanced Supercomputing Division at Ames Research Center, and XSEDE TACC (TG-AST180008). 
DC was partially supported by NASA (grants 80NSSC18K1218, 80NSSC20K1273, and 80NSSC18K1726) and by NSF (grants AST-1714658, AST-2009326, AST-1909778, PHY-1748958, and PHY-2010240). 
\appendix
\section{Details of the analytic calculations}\label{app:disper}
At first let us recall the Amp\`{e}re-Maxwell equation: $\nabla \times {\bf B}=\frac{4\pi}{c} {\bf J} +\frac{1}{c} \frac{\partial}{\partial t}  {\bf E}$ and the Maxwell-Faraday equation: $\nabla \times {\bf E}=-\frac{1}{c} \frac{\partial}{\partial t}  {\bf B}$. 
%In the absence of current source (denoted by ${\bf J}$), these equations determine propagation of the electromagnetic waves.
Here we will show that a non-zero ${\bf J}$ that comes from unbalanced perturbed current in the plasma generates waves, which can grow/damp/oscillate depending on the modes.

Initially, the bulk speed ($v_{\rm e}$) of background electrons (Equation \ref{eq:veloe}) balances the CR current, i.e., the total $J=0$. 
Suppose plane-wave perturbations are imposed on the background electromagnetic fields, which result in density and velocity fluctuations in the background ions and electrons.
Denoting the first-order perturbations with the subscript $_1$, the total current density at $t>0$, in the CR + plasma composite system:
\begin{eqnarray} \label{eq:Jtot}
{\bf J} &= & \left[s_{\rm cr}\,e\,n_{\rm cr}\,{\bf v}_{\rm d}\right] +  \left[e\left(n_{\rm i} + n_{\rm 1i}\right){\bf v}_{\rm 1i}\right] +\left[- e\left(n_{\rm e} + n_{\rm 1e}\right)\left({\bf v}_{\rm e}+{\bf v}_{\rm 1e}\right)\right]= e n_{\rm i}({\bf v}_{\rm 1i}-{\bf v}_{\rm 1e}) - e\,n_{\rm 1e}{\bf v}_{\rm e} - s_{\rm cr}\,e\,n_{\rm cr}{\bf v}_{\rm 1e} . 
\end{eqnarray}
Velocity and density perturbations introduced in Equation \ref{eq:Jtot} are obtained as follows. As the perturbations on EM field are modulated with $\exp[j(k\,x-\omega\,t)]$ (where ${\bf k}=k\,\hat{\bf x}$ is the propagation vector and $\omega$ is the angular frequency), linearization of the Lorentz force equation (Equation \ref{eq:nonrel_motion}) gives
\begin{eqnarray}\label{eq:velo1x}
v_{\rm 1x\alpha} & = & \frac{jq_{\rm \alpha}}{m_{\rm \alpha}\omega} E_{\rm 1x}\\
\label{eq:velo1y}
v_{\rm 1y\alpha} & = & \frac{jq_{\rm \alpha}}{m_{\rm \alpha}\omega}\left[\frac{1-v_{\rm 0}\,k/\omega}{1-\left(\omega_{\rm c\alpha}/\omega\right)^2}\left(E_{\rm 1y}+j \frac{\omega_{\rm c\alpha}}{\omega} E_{\rm 1z}\right)\right]\\
\label{eq:velo1z}
v_{\rm 1z\alpha} & = & \frac{jq_{\rm \alpha}}{m_{\rm \alpha}\omega}\left[\frac{1-v_{\rm 0}\,k/\omega}{1-\left(\omega_{\rm c\alpha}/\omega\right)^2}\left(-j \frac{\omega_{\rm c\alpha}}{\omega}E_{\rm 1y}+ E_{\rm 1z}\right)\right] \ ,
\end{eqnarray}
where we have used the linearized Maxwell-Faraday equation (given below) to substitute the ${\bf B}$-field:
\begin{equation}\label{eq:MFeq}
B_{\rm 1y} = -\frac{k\,c}{\omega}E_{\rm 1z}\, , \ {\rm and}\ B_{\rm 1z} =  \frac{k\,c}{\omega} E_{\rm 1y}
\end{equation}
In Equations \ref{eq:velo1y} and \ref{eq:velo1z}, $\omega_{\rm c\alpha} = q_{\rm \alpha} B_{\rm 0}/m_{\rm \alpha} c$ is the cyclotron frequency and $v_{\rm 0}=s_{\rm cr} v_{\rm e} \neq 0$ only for electrons ($v_{\rm e} = |{\bf v}_{\rm e}$; Equation \ref{eq:veloe}).
The density fluctuations can be obtained from the ion and electron mass continuity equations, which give $n_{\rm 1i}=n_{\rm i} k\, v_{\rm 1xi}/\omega$ and $n_{\rm 1e}=n_{\rm e} k\, v_{\rm 1xe}/(\omega-k\,v_{\rm e}s_{\rm cr})$ respectively.
Substituting Equations \ref{eq:velo1x} - \ref{eq:velo1z} in Equation \ref{eq:Jtot} and neglecting higher order (more than one) terms of $\omega/\omega_{\rm ci}$ and $\omega/\omega_{\rm ce}$ (as our regime of interest $\omega \ll \omega_{\rm ci}$), we obtain,
\begin{eqnarray}\label{eq:jtotdetails} 
\label{eq:J1x}
J_{\rm 1x} &= &\frac{j\,E_{\rm 1x}}{4\pi\omega}\left[\omega_{\rm pi}^2+\omega_{\rm pe}^2\left(1+\frac{s_{\rm cr}v_{\rm e}k}{\omega-s_{\rm cr}v_{\rm e}k} +s_{\rm cr} \frac{n_{\rm cr}}{n_{\rm i}}\right)\right] \\
\label{eq:J1y}
J_{\rm 1y} &=-&\frac{j\,c\,\omega_{\rm pi}}{4\pi v_{\rm A0}}
\left[E_{\rm 1y}\left\{\frac{\omega}{\omega_{\rm ci}}+\left(\frac{\omega}{\omega_{\rm ce}}-s_{\rm cr}\frac{v_{\rm e}k }{\omega_{\rm ce}}\right)\left(1+s_{\rm cr}\frac{n_{\rm cr}}{n_{\rm i}}\right)\right\} 
+ j\,E_{\rm 1z}\left\{s_{\rm cr}\frac{v_{\rm e}k }{\omega}\left(1+s_{\rm cr}\frac{n_{\rm cr}}{n_{\rm i}}\right)-s_{\rm cr}\frac{n_{\rm cr}}{n_{\rm i}}\right\}\right]\\
%
%\label{eq:jtotdetails3}
\label{eq:J1z}
J_{\rm 1z} &=-&\frac{j\,c\,\omega_{\rm pi}}{4\pi v_{\rm A0}}
\left[- j\,E_{\rm 1y}\left\{s_{\rm cr}\frac{v_{\rm e}k }{\omega}\left(1+s_{\rm cr}\frac{n_{\rm cr}}{n_{\rm i}}\right)-s_{\rm cr}\frac{n_{\rm cr}}{n_{\rm i}}\right\}
+E_{\rm 1Z}\left\{\frac{\omega}{\omega_{\rm ci}}+\left(\frac{\omega}{\omega_{\rm ce}}-s_{\rm cr}\frac{v_{\rm e}k }{\omega_{\rm ce}}\right)\left(1+s_{\rm cr}\frac{n_{\rm cr}}{n_{\rm i}}\right)\right\} 
\right]
\end{eqnarray}
Here we have taken $B_{\rm 0}/\left(4\pi\,m_{\rm i}\, n_{\rm 0}\right)^{1/2}$ as the Alfv\'{e}n speed $v_{\rm A0}$ (since $m_{\rm i}\gg m_{\rm e}$, we can take $v_{\rm A0i} = v_{\rm A0}(1+m_{\rm e}/ m_{\rm i})^{1/2}\simeq v_{\rm A0})$. $\omega_{\rm ci,e}= |e B_{\rm 0}/m_{\rm i,e}c|$ is cyclotron frequency, $\omega_{\rm pi,e}=\left(4\pi\,n_{\rm 0}e^2/m_{\rm i,e}\right)^{1/2}$ is the plasma frequency for ions/electrons. 
%, and $v_{\rm A0}= B_{\rm 0}/\left(4\pi\,m_{\rm i}\, n_{\rm 0}\right)^{1/2}$ is the Alfv\'{e}n speed for ions (taking $n_{\rm i}\approx n_{\rm e}=n_{\rm 0}$).
Equations (\ref{eq:J1x}) - (\ref{eq:J1z}) show that perturbed current density is non-zero, which act as a source in the Amp\`{e}re's-Maxwell equation. Since we assume $n_{\rm cr} \ll n_{\rm e}$, the transverse components of the current are simplified to 
$J_{\rm 1y}, J_{\rm 1z} \approx (\omega^2/k)\,(-s_{\rm cr}B_{\rm 1y}, s_{\rm cr}\,B_{\rm 1z})c/(4\pi\,v_{\rm A0}^2)$, indicating a direct dependency on the transverse magnetic fields, i.e., a tiny perturbation in the magnetic field can increase
the current, which further amplifies the magnetic field and so on.
\subsection{Dispersion relation}
Substituting Equations \ref{eq:J1x}-\ref{eq:J1z} in the Amp\`{e}re-Maxwell equation, and combining the Maxwell-Faraday equations:
\begin{eqnarray}\label{eq:alphabeta}
 \left[
\begin{array}{c c c}
h_{\rm 1} & 0 & 0 \\ 
0 & h_{\rm 2}  & -j\, h_{\rm 3}  \\
0 & j\, h_{\rm 3}  & h_{\rm 2} 
\end{array}     \right] \cdot \left[
\begin{array}{c}
E_{\rm 1x} \\
E_{\rm 1y}\\
E_{\rm 1z}\\
\end{array}     \right] &=& 
\left[
\begin{array}{c}
0 \\
0\\
0
\end{array}     \right]\, , \, {\rm where}
\end{eqnarray}
\begin{equation} %\label{eq:det_h1}
h_{\rm 1} = 1 - \frac{1}{\omega^2}\left[\omega_{\rm pi}^2+\omega_{\rm pe}^2\left(1+\frac{s_{\rm cr}v_{\rm e}k}{\omega-s_{\rm cr}v_{\rm e}k} +s_{\rm cr} \frac{n_{\rm cr}}{n_{\rm i}}\right)\right],\nonumber 
\end{equation}
\begin{equation}
% \label{eq:det_h2}
h_{\rm 2} = k^2- \frac{\omega^2}{c^2}  - \frac{\omega \omega_{\rm pi}}{c\,v_{\rm A0}}
\left[\frac{\omega}{\omega_{\rm ci}}+\left\{\frac{\omega}{\omega_{\rm ce}}-s_{\rm cr}\frac{v_{\rm e}k }{\omega_{\rm ce}}\right\}\left(1+s_{\rm cr}\frac{n_{\rm cr}}{n_{\rm i}}\right)\right],\, {\rm and}\, \
%\label{eq:det_h3}
h_{\rm 3} = \frac{\omega \omega_{\rm pi}}{c\,v_{\rm A0}}\left[s_{\rm cr}\frac{v_{\rm e}k }{\omega}\left(1+s_{\rm cr}\frac{n_{\rm cr}}{n_{\rm i}}\right)-s_{\rm cr}\frac{n_{\rm cr}}{n_{\rm i}}\right]
\end{equation}
Equation \ref{eq:alphabeta} gives two distinct solutions:

\noindent {\bf Solution A: $E_{\rm 1\,y,z}= 0$.} In this case, if $v_{\rm e}=0$, then $\omega \approx (\omega_{\rm pi}^2+\omega_{\rm pe}^2)^{1/2}$, where $\omega_{\rm pi,e}=\left(4\pi\,n_{\rm 0}e^2/m_{\rm i,e}\right)^{1/2}$ is the plasma frequency for ions/electrons. This represents plasma oscillations.

\noindent {\bf Solution B:} $h_{\rm 2}=\pm h_{\rm 3}$, we find a quadratic equation of $\omega$: $
\omega^2 \mathcal{A} - \omega  \mathcal{B} - \mathcal{C} = 0$ which provides the dispersion relation in the following form.
\begin{equation} \label{eqaa:disp_full1}
\omega= \frac{\mathcal{B} + \left[\mathcal{B}^2+ 4\,\mathcal{A}\,\mathcal{C}\right]^{1/2}}{2  \mathcal{A}}  \ ,{\rm where}
\end{equation}
\begin{equation}\label{eqapp:disp_full2}
\mathcal{A} = \left[\left(\frac{v_{\rm A0}}{c}\right)^2+1+\frac{\omega_{\rm ci}}{\omega_{\rm ce}}\left(1+s_{\rm cr}\frac{n_{\rm cr}}{n_{\rm i}}\right)\right],\,
\mathcal{B} = \omega_{\rm ci}
\left[
s_{\rm cr}\frac{v_{\rm e}k}{\omega_{\rm ce}}(1+s_{\rm cr}\frac{n_{\rm cr}}{n_{\rm i}})
\pm s_{\rm cr}\frac{n_{\rm cr}}{n_{\rm i}}
\right],\,
\mathcal{C} = k^2\,v_{\rm A0}^2\left[1\mp s_{\rm cr} \frac{k_{\rm u}}{k}\, \left(1+s_{\rm cr}\frac{n_{\rm cr}}{n_{\rm i}}\right)\right]
\end{equation}
where we have introduced a parameter $k_{\rm u} = \omega_{\rm pi}\, |v_{\rm e}|/c\, v_{\rm A0}$. Using $\omega_{\rm ci} = (v_{\rm A0}/c)\,\omega_{\rm pi}$, we obtain a simplified expression of $\mathcal{B}$:
$\mathcal{B} = v_{\rm A0}\,k_{\rm u}\left[s_{\rm cr}\frac{k}{k_{\rm u}}\frac{m_{\rm e}}{m_{\rm i}}\frac{v_{\rm e}}{v_{\rm A0}}\left(1+s_{\rm cr}\,\frac{n_{\rm cr}}{n_{\rm i}}\right) \pm s_{\rm cr}\frac{v_{\rm A0}}{v_{\rm d}}\right]$ which is simplified to $\mathcal{B} \approx \pm s_{\rm cr}\,v_{\rm A0}^2\,k_{\rm u}/v_{\rm d}$. It can be shown that when $v_{\rm A0} \ll v_{\rm d}$, the term under square-root in Equation \ref{eqaa:disp_full1} mostly depends on $4 \mathcal{A}\,\mathcal{C} $, i.e., the  square-root term can be a complex number depending on the ratio $k/k_{\rm u}$. Using these assumptions, Equation \ref{eq:disp} is obtained. Note that, if these conditions are not satisfied, then one can still obtain growing modes, however, the wavelength of the fastest growing mode and the growth rate can deviate from \cite{bell04}'s prediction due to contribution of $\mathcal{B}$.% to the square-root term of Equation \ref{eqaa:disp_full1}.

Equations \ref{eq:MFeq} and \ref{eq:alphabeta} suggest that the transverse B-field, ${\bf B}_{\rm \perp}=B_{\rm 1y} {\bf \hat{y}}+B_{\rm 1z}{\bf \hat{z}}\propto [\hat{y} \exp(\pm j\pi/2)+\hat{z}]$, i.e., $\Delta \phi=\pm \pi/2$ (Equation \ref{eq:disp_0}). To find the phase difference between $B_{\rm y}$ and $B_{\rm z}$ for a given mode, $k$, from our simulation, we have used Equation \ref{eq:polangle}, where
\begin{equation}\label{eq:appstokes}
 Q(k)  = \left[\tilde{B}_{\rm y}(k) \tilde{B}_{\rm y}^*(k) - \tilde{B}_{\rm z}(k) \tilde{B}_{\rm z}^*(k)\right],\,
 U(k) =  \left[\tilde{B}_{\rm y}(k)\tilde{B}_{\rm z}^*(k) + \tilde{B}_{\rm y}^*(k) \tilde{B}_{\rm z}(k)\right],\,
 V(k) = j \left[\tilde{B}_{\rm y}(k)\tilde{B}_{\rm z}^*(k) - \tilde{B}_{\rm y}^*(k) \tilde{B}_{\rm z}(k)\right]
\end{equation}
Here $\tilde{B}_{\rm y,z}(k)$ are the Fourier transform of $B_{\rm y,z}(x)$ along the $x$ axis (the superscript `$*$' denotes the complex conjugate).

\subsection{Back-reaction and saturation}\label{app:sat}
The above derivation does not include the back-reaction from the plasma caused by the growing waves. 
In later times ($t\gg \gamma_{\rm fast}^{-1}$), the force due to the term  e.g. ${\bf  J}\times {\bf B}/c$ can affect the momentum of CRs and plasma. 
The unstable waves cannot grow for an indefinite amount of time and saturate. 
Below we extend our linear analysis to predict the saturation which is based on the fundamental fact that the net momentum deposited by CRs goes into thermal background through amplified field. 
Note that the saturation is a non-linear process and numerical simulation can provide a better result and therefore our prediction should be treated as an approximated solution.

Let us recall a more general form of the momentum equation of the plasma:
\begin{eqnarray}\label{eqapp:mb}
 \frac{\partial}{\partial t}\left[n_{\rm 0}(m_{\rm i}{\bf v_{\rm i}}+m_{\rm e}{\bf v_{\rm e}})\right] & \approx & \frac{1}{c}\left( {\bf J} \times {\bf B}\right)-  \nabla (P_{\rm i}+P_{\rm e}) .
\end{eqnarray}
Here $P_{\rm i,e}$ is ion/electron pressure in the plasma. 
We shall take into account two terms in the right side of Equation \ref{eqapp:mb} one by one as done to obtain an approximated solution.
Firstly assuming that the second term in the right hand side (RHS) is much smaller than the first term, we obtain the velocity of plasma ions/electrons:
\begin{equation}\label{eq:pacc}
v_{\rm 1xi,e} \approx \frac{v_{\rm A0}^2}{v_{\rm d}} \left(\frac{B_{\rm\perp}}{B_{\rm 0}}\right)^2,\ {\rm and}\ v_{\rm 1y/zi,e} \approx v_{\rm A0} \left(\frac{B_{\rm\perp}}{B_{\rm 0}}\right) .
\end{equation}
Since we start with $v_{\rm A0}\ll v_{\rm d}$, $B_{\rm \perp} =0$, $v_{\rm 1xi,e} \rightarrow 0$. 
With time, the growing $B_{\rm \perp}$ results in plasma acceleration. 
Therefore, the plasma ions that were initially treated stationary with respect to the lab frame also start drifting along $x$-direction. 
Whereas, the equal raise in transverse velocity components mainly contributes to increase velocity dispersion of the plasma, and raises the plasma temperature. 
Assuming the initial thermal energy per particle in the plasma $\approx  m\,a_{\rm 0}^2/2\sim k_{\rm B} T_{\rm 0}/2$ ($T_{\rm 0}$ as the initial temperature), the final temperature of the plasma is expected to be
\begin{equation}\label{eq:Theating}
T \sim T_{\rm 0}\left[1+ \left(\frac{v_{\rm A0}}{a_{\rm 0}}\right)^2 \left(\frac{B_{\rm\perp}}{B_{\rm 0}}\right)^2\right]% =T_{\rm 0}\left[1+\frac{2}{\beta}\,\left(\frac{B_{\rm \perp}}{B_{\rm 0}}\right)^2\right].
\end{equation}
%Here $\beta$ is the ratio of the gas pressure to the magnetic pressure. 
Therefore, a larger $B_{\rm \perp}/B_{\rm 0}$ implies an intense heating effect. 
The second term in RHS of Equation \ref{eqapp:mb} that represents the loss in momentum due to plasma heating is  calculated by using Equation \ref{eq:Theating}:
\begin{eqnarray}
\int\, dt\, \nabla (P_{\rm i} + P_{\rm e})|_{\rm x} \approx \int dt\,  {\rm Re}\left[\frac{j\,k^2 }{\omega}\right] (n_{\rm 0}\,m_{\rm i}\,a_{\rm i}^2\,v_{\rm 1ix}) & \approx &\int dt\, \frac{k}{v_{\rm A0}}\left[ n_{\rm 0}\,m_{\rm i}\left\{a_{\rm i0}^2+v_{\rm A0}^2\left(\frac{B_{\rm \perp}}{B_{\rm 0}}\right)^2\right\}\frac{v_{\rm A0}^2}{v_{\rm d}}\left(\frac{B_{\rm \perp}}{B_{\rm 0}}\right)^2\right] %\nonumber\\
%& =&  \frac{n_{\rm 0}m_{\rm i}}{4}\frac{v_{\rm A0}^2}{v_{\rm d}}\left[2\left(\frac{a_{\rm i0}^2}{v_{\rm A0}}\right)^2\,\left(\frac{B_{\rm \perp}}{B_{\rm 0}}\right)^2+\left(\frac{B_{\rm \perp}}{B_{\rm 0}}\right)^4\right]
\end{eqnarray}
Now considering that the net momentum deposited by CRs goes into thermal background, the time integration of x-component of Equation \ref{eqapp:mb} yields,
\begin{eqnarray}\label{eqapp:mbt}
\left[n_{\rm cr}p_{\rm cr,x}\right]|_{\rm t=0}^{t} & \approx &  -n_{\rm 0}\,m_{\rm i}\frac{v_{\rm A0}^2}{v_{\rm d}}\left(\frac{B_{\rm \perp}}{B_{\rm 0}}\right)^2  - \int\, dt\, \nabla (P_{\rm i} + P_{\rm e})|_{\rm x}
\end{eqnarray}
LHS: At $t=0$, $n_{\rm cr}p_{\rm cr,x}=n_{\rm cr}(\gamma_{\rm bst}\,v_{\rm bst}\,\,E_{\rm cr}^{\prime}/c^2)$ (Equation \ref{eq:pcrlab}). 
We further assume that in the final stage, the drift velocity of CRs $ \tilde{v}_{\rm d} \approx v_{\rm A}$ (as observed in the simulation). 
This gives $\tilde{v}_{\rm bst}\approx v_{\rm A}$ and $\tilde{\gamma}_{\rm bst}\simeq 1$, i.e., $n_{\rm cr}p_{\rm cr,x}\approx n_{\rm cr}(v_{\rm A}\,E_{\rm cr}^{\prime}/c^2)$. We finally obtain:
\begin{equation}\label{eq:pqrs}
\mathcal{P}\left(\frac{B_{\rm \perp}}{B_{\rm 0}}\right)^4+\mathcal{Q}\,\left(\mathcal{\frac{B_{\rm \perp}}{B_{\rm 0}}}\right)^2 + \mathcal{R}\left(\mathcal{\frac{B_{\rm \perp}}{B_{\rm 0}}}\right)  - \mathcal{S}= 0 \, ,
\end{equation}
where $\mathcal{P}=1$, $\mathcal{Q}= \left[2\left(a_{\rm i0}/v_{\rm A0}\right)^2+4\right]$, $\mathcal{R}=\left[4\frac{n_{\rm cr}}{n_{\rm 0}} \frac{E^{\prime}_{\rm cr}}{m_{\rm i}c^2}\frac{v_{\rm d}}{v_{\rm A0}}\right]$ and  $\mathcal{S}=\left[4\frac{n_{\rm cr}}{n_{\rm 0}}\frac{\gamma_{\rm bst} v_{\rm bst}E^{\prime}_{\rm cr}/c^2}{m_{\rm i}c}\frac{v_{\rm d}\,c}{v_{\rm A0}^2}\right]$. 
The above equation can be solve numerically and the approximated solution is %$B_{\perp}/B_{\rm 0}\approx [(b^2+4\xi)^{1/2}-b]^{1/2}$, 
\begin{equation}\label{eq:approBsat_l}
    \frac{B_{\perp}}{B_{\rm 0}} \approx [(b^2+4\xi)^{1/2}-b]^{1/2} \, , {\rm  where}\ \, b=2+\left(\frac{a_{\rm i0}}{v_{\rm A0}}\right)^2\ {\rm and}\ \, \xi=\left[\frac{n_{\rm cr}}{n_{\rm 0}}\frac{\gamma_{\rm bst} v_{\rm bst}E^{\prime}_{\rm cr}/c^2}{m_{\rm i}c}\frac{v_{\rm d}\,c}{v_{\rm A0}^2}\right]^{1/2} .
\end{equation}
%where $b=2+(a_{\rm i0}/v_{\rm A0})^2$ and $\xi=\left[\frac{n_{\rm cr}}{n_{\rm 0}}\frac{\gamma_{\rm bst} v_{\rm bst}E^{\prime}_{\rm cr}/c^2}{m_{\rm i}c}\frac{v_{\rm d}\,c}{v_{\rm A0}^2}\right]^{1/2}$. 
If we neglect the heating losses and take $a_{\rm i0}\ll v_{\rm A}$ (i.e., the term $\mathcal{P}$ is absent and $\mathcal{Q}\rightarrow 4$: cold plasma), then Equation \ref{eq:pqrs} gives $B_{\perp}/B_{\rm 0}\approx (\mathcal{S}/\mathcal{Q})^{1/2}=\xi^{1/2}$, which is identical to Equation \ref{eq:bellsat}. 
These two possible solutions of $B_{\rm \perp}/B_{\rm 0}$ are referred as the lower and upper limits of $B_{\perp}/B_{\rm 0}$ and are shown by the cyan lines in Figure \ref{fig:allin1}.
%%%%%%%%%%%%%%%%%%%%%%%%%%%%%%%%%%%%%%%%%%%%%%%%%%%%%%%%
\section{How large $\xi$ should be chosen}\label{appsec:xi_experiment}
%%%%%%%%%%%%%%%%%%%%%%%%%%%%%%%%%%%%%%%%%%%%%%%%%%%%%%%%
\begin{figure}
\centering
\includegraphics[width=0.37\textwidth]{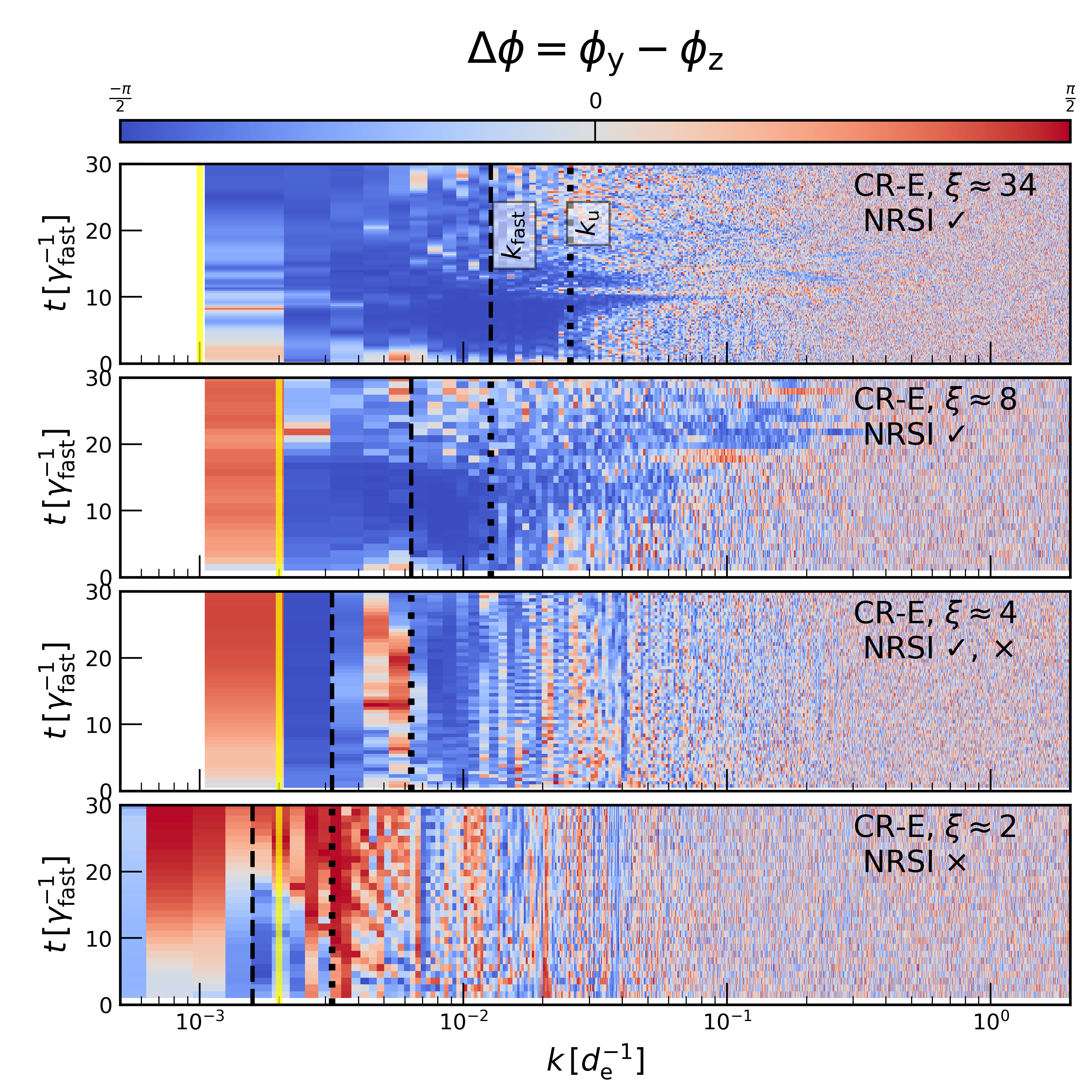}
\caption{Phase difference $\Delta \phi$ for different modes as a function of time with descending $\xi$ (upper to lower panels). The yellow lines show $k=1/R_{\rm L}$. For the upper two panels, the results are consistent with linear theory. For a smaller $\xi$, the NRSI becomes sub-dominated and the linear theory outlined here is not applicable.}\label{fig:diffXi}
\end{figure}
A general assumption in the NRSI is that $\xi \gg 1$ (Equation \ref{eq:nrlimit}). Here we explore how large the value of $\xi$ must be chosen to apply the standard theory of NRSI safely. From section \ref{sec:analytics}, we recall that the growing modes in the CR-E case have negative helicity. 
Following the results presented in \S\ref{subsec:maximally}, we check how $\Delta\phi(k)$ changes as a function of $\xi$ by altering $v_{\rm A0}$ and $n_{\rm cr}$ (other parameters similar to Run B in Table \ref{tab:simpara}). 
Figure \ref{fig:diffXi} indicates that the dominating modes have $\Delta \phi \approx -\pi/2$ (blue regions) when $\xi \gtrsim 4$, i.e., below $\xi \approx 4$, the NRSI and the RSI blend into each other.

%\bibliography{sample63,Total}{}

\bibliography{Total}{}
\bibliographystyle{aasjournal}

\end{document}